\documentclass[12pt,a4paper]{article}
\usepackage{epsf}
\def\d{\, {\rm d}}
\def\LPP{Li\v{s}\v{c}i\'{c}/Petrofer probe}
\newcounter{myfig}
\def\Fig#1{\refstepcounter{myfig}\label{#1}}
\author{Sa\v{s}a Singer%
\thanks{Department of Mathematics, Faculty of Science,
	Bijeni\v{c}ka cesta 30, HR--10000 Zagreb, Croatia
	(e-mail: singer@math.hr)}}
\title{Sensitivity of the Heat Transfer Coefficient Calculation}

\begin{document}
%
%
%
%
\thispagestyle{plain}
%
%
\begin{center}
{\Large Sa\v{s}a Singer}\large
\footnote{Department of Mathematics, Faculty of Science,
	Bijeni\v{c}ka cesta 30, HR--10000 Zagreb, Croatia
	(e-mail: singer@math.hr)}
\end{center}

\vspace*{-3pt}
%
%
\begin{center}
{\Large Sensitivity of the Heat Transfer Coefficient Calculation}
\end{center}

\vspace*{6pt}
%
%
\centerline{Dedicated to the memory of prof.\ Mladen Rogina, 1957--2013}

\vspace*{12pt}
%
%
%
%
\begin{abstract}
	The purpose of the \LPP\ is to determine the cooling intensity
during liquid quenching in laboratory and workshop environments.
The surface heat transfer coefficient is calculated by the one-dimensional
finite volume method from the smoothed temperature curve, measured at
a near-surface point in the probe.
	Smoothed reference temperature curves for oil and water, based on
measurements made by the probe, are used in a series of numerical experiments
to investigate the sensitivity of the calculation with respect to various
input parameters of the problem, such as: thermal properties of the material,
the near-surface thermocouple position, and the diameter of the probe.
These results are relevant for other liquid quenchants, at least
qualitatively, if not quantitatively.
	A similar type of analysis is conducted with respect to the variation
of numerical simulation parameters in the finite volume method---the choice
of space and time steps for problem discretization. When the input curves
are sufficiently smooth, the method itself is very reliable.
\end{abstract}

\vspace*{6pt}

\vspace*{12pt}
\noindent
{\small
%
}

\newpage
%
%
%
%
\hrule
\begin{center}
{\bf Nomenclature}
\end{center}
\hrule

%
%
\vspace*{6pt}

\vspace*{6pt}
$t$ = time, ${\rm s}$

\vspace*{6pt}
$t_{\rm final}$ = final time for measured temperatures and numerical
	simulation, ${\rm s}$

\vspace*{6pt}
$t_i$ = discrete time level $i$ in time discretization for numerical
	simulation, ${\rm s}$

\vspace*{6pt}
$\tau_i$ = time step between successive time levels in time discretization,
	${\rm s}$

\vspace*{6pt}
$\tau$ = finest time step, used at the beginning of the simulation,
	${\rm s}$

\vspace*{6pt}
$r$ = radial space coordinate, ${\rm m}$ (or ${\rm mm}$)

\vspace*{6pt}
$D$ = diameter of the probe, ${\rm m}$ (or ${\rm mm}$)

\vspace*{6pt}
$R$ = radius of the probe, ${\rm m}$ (or ${\rm mm}$)

\vspace*{6pt}
$d_n$ = depth of the near-surface thermocouple in the probe, ${\rm m}$
	(or ${\rm mm}$)

\vspace*{6pt}
$h$ = space step in space discretization, ${\rm m}$ (or ${\rm mm}$)

\vspace*{6pt}
$N_{\rm ext}$ = number of solution extension intervals in space discretization

\vspace*{6pt}
$T$ = temperature, ${\rm K}$ or ${\rm {}^\circ C}$

\vspace*{6pt}
$T_n$ = near-surface temperature, ${\rm K}$ or ${\rm {}^\circ C}$

\vspace*{6pt}
$T_x$ = external temperature of the quenchant, ${\rm K}$ or ${\rm {}^\circ C}$

\vspace*{6pt}
$T_s$ = surface temperature, ${\rm K}$ or ${\rm {}^\circ C}$

\vspace*{6pt}
$T_c$ = core (center) temperature, ${\rm K}$ or ${\rm {}^\circ C}$

\vspace*{6pt}
$\lambda$ = thermal conductivity, ${\rm W / (m \cdot K) }$

\vspace*{6pt}
$c$ = specific heat, ${\rm J / (kg \cdot K) }$

\vspace*{6pt}
$\rho$ = density (mass density), ${\rm kg / m^3 }$

\vspace*{6pt}
$\gamma = c \times \rho$ = volumetric specific heat,
	${\rm J / (m^3 \cdot K) }$

\vspace*{6pt}
$q$ = heat flux density, ${\rm W / m^2}$

\vspace*{6pt}
$q_s$ = surface heat flux density, ${\rm W / m^2}$

\vspace*{6pt}
$\alpha$ = heat transfer coefficient (HTC) at the surface,
	${\rm W / (m^2 \cdot K) }$

\vspace*{6pt}
$\vec{n}$ = unit outer vector perpendicular to the surface

\vspace*{12pt}
\hrule

\newpage
%
%
%
%
%
%
\section{Introduction}
%
%
	Quenching probes are standard tools for measuring and evaluating
the cooling intensity of various liquid and gas quenchants. They are equipped
with one or more thermocouples, that measure and record the temperature
curves at selected positions in the probe throughout the whole quenching
period of time. The cooling rate, i.e., the time derivative of the temperature
at a particular point, can be estimated from the measured cooling curve by
approximate numerical differentiation. These two curves (in time) are
the basis of all cooling curve tests
\cite{Totten-Dossett-Kobasko-2013,Liscic-Singer-2013}.
Because the calculation of all relevant data is relatively simple, for many
years, such tests have been the most useful means for a quick characterization
of quenchants, without estimating the actual cooling conditions at
the surface---where the heat extraction occurs.
In modern times, when the computing power is much greater, standard cooling
curve tests are becoming inadequate for accurate simulation and control
of the quenching conditions in industrial applications.

	The cooling ability of a quenchant is characterized by the heat flux
between the hot metal surface and the quenchant. An appropriate model for
this heat extraction is Newton's law of cooling \cite{Hernandez-Morales-2013},
combined with Fourier's law of heat conduction
\begin{equation}
	q_s = - \lambda \, \frac{\partial T}{\partial \vec{n}} =
		\alpha \left( T_s - T_x \right),
\label{Eq.1}
\end{equation}
where:
\begin{itemize}
  \item[]
	$q_s$ = surface heat flux density, ${\rm W / m^2}$,
	considered positive when the probe cools,
  \item[]
	$\vec{n}$ = unit outer vector perpendicular to the surface,
  \item[]
	$T_s$ = surface temperature at a particular point,
	${\rm K}$ or ${\rm {}^\circ C}$,
  \item[]
	$T_x$ = external temperature of the quenchant,
	${\rm K}$ or ${\rm {}^\circ C}$, and
  \item[]
	$\alpha$ = heat transfer coefficient (HTC) at the surface,
	${\rm W / (m^2 \cdot K) }$.
\end{itemize}
The HTC is regarded as the measure of the cooling power of a quenchant,
and Eq~1 is used as the boundary condition in numerical simulation of
the quenching process, involving solution of the heat conduction equation
to predict the hardness of steels (see, for example,
\cite{Smoljan-1998,Smoljan-2002,Liscic-Singer-Smoljan-2010}).

	The reference HTC curves for various quenchants and quenching
conditions can be estimated by solving the inverse heat conduction problem
(IHCP), where the boundary condition is reconstructed from available
input data---one or more measured cooling curves gathered by a particular
probe. An overview of different methods for calculating the HTC is given in
\cite{Totten-Dossett-Kobasko-2013}. However, there are several practical
problems common to all methods.

	In all experiments, temperatures are measured only at a few points
in the probe. It is well known that intensive quenching is not a regular
process, and quenching conditions, i.e., the actual HTC values, may be
very different at various places on the surface
\cite{Totten-Dossett-Kobasko-2013}. As a consequence, regardless of
the method, all calculated HTCs are some approximations and should be
interpreted and used with some caution. The only way to increase
the reliability of estimated HTCs is by repeating the measurements in similar
quenching conditions.

	The calculation of the HTC from measured temperature curves is
an ill-posed problem, meaning that the solution is not well-determined by
the input data, and it is very sensitive to measurement errors and small
perturbations of the data \cite{Hernandez-Morales-2013,%
Tikhonov-Arsenin-1977,Beck-Blackwell-StClair-1985}.
Therefore, some form of model transformation and regularization is necessary
to obtain a useful solution.

	Measured cooling curves are the main input data, and any noise in
these curves is very likely to be magnified in the calculated HTC curves.
So, the most common initial regularization step is to apply some smoothing
procedure to reduce or eliminate the so-called random noise in these data.
A general approach to both smoothing and approximation of measured
temperatures is described in \cite{Liscic-Singer-2013,Liscic-Singer-2014}.
Earlier, the effect of different smoothing techniques was analyzed in
\cite{Felde-et-al-2002}.

	Many of the results published in the literature about the HTC are
based on relatively small probes, that have only one thermocouple located
at the center of the probe (like the ISO 9950 probe). The so-called damping
effect frequently results in underestimated peak HTC values. An agreement of
the results computed by various methods for the HTC calculation is observed,
as in \cite{Felde-et-al-2009}, only when the temperature is measured
very near or at the surface of the probe.
Finally, the HTC values for small probes are much larger then those for
larger probes. From the discussion in \cite{Liscic-Singer-2013}, it follows
that the results obtained with small probes have a limited applicability for
quenching of much larger pieces in industrial processes.

	Apart from that, relatively little is known about the sensitivity
of the HTC calculation with respect to all other parameters of the problem,
especially when larger probes are used, so that all simple methods for
the HTC calculation are not feasible.

	The aim of this paper is to fill that gap in the literature and
provide some information about the sensitivity of the HTC calculation
procedure, and, hopefully, of the HTC itself, with respect to some of
the variables involved in the quenching process.
The power of numerical simulation is used to perform a number of
``experiments'' that would be very expensive or even impossible to do in
real life, with real probes in actual quenching conditions.
To increase the practical value of the results, the initial
cooling curves are based on the actual data recorded by the \LPP---a large
probe designed for characterization of industrial quenching processes
\cite{Liscic-Singer-2013,Liscic-Singer-2014}.

	It must be said immediately that this approach is prone to systematic
errors in the model and in the software implementation, but it is believed
that such errors, if they exist, affect all the experiments in a similar way,
so that the results of the sensitivity analysis (with small perturbations)
are still reasonably accurate. To verify that, the same type of tests is
applied to the actual HTC calculation algorithm.

	A brief description of the probe and the HTC calculation is given in
the next section, followed by some general sensitivity concerns that influence
the design and the scope of all experiments. The results of individual
experiments are presented in subsequent sections.

	Even though the experiments are based on the data gathered by
a particular probe, in only two liquid quenchants, and conducted with
a particular algorithm for solving the IHCP, the results presented here
should be relevant for other quenchants and other algorithms, that are based
on similar principles.

%
%
\section{Heat transfer coefficient calculation}
%
%
	Complete descriptions of the \LPP\ and the HTC calculation procedure
are given in \cite{Liscic-Singer-2013,Liscic-Singer-2014}, and will be
only briefly summarized here.

	The probe is a $200 \, {\rm mm}$ long cylinder with the diameter
$D = 50 \, {\rm mm}$, made of an austenitic nickel--chromium--iron superalloy
(Inconel 600\footnote{Inconel alloy 600$^{\rm TM}$ is a trademark of Special
Metals Corporation, USA.}, UNS N06600), that does not have phase
transformations with the latent heat involved.

	The near-surface temperature, denoted by $T_n$ (${\rm {}^\circ C}$),
is measured in time $t$ (${\rm s}$) at the depth $d_n = 1 \, {\rm mm}$ below
the surface, with the highest frequency of $50$ measurements per second.
Two additional temperatures curves are measured in the probe, one at the depth
of $4.5 \, {\rm mm}$ below the surface, and the other one at the center
(core). The external quenchant temperature $T_x$ is also known, usually as
a constant value at the beginning of the quenching test.

	To compensate for the lack of input data, the whole calculation is
based on a simplified one-dimensional (1D) model of heat conduction.
The probe is considered as a long, radially symmetrical cylinder, with
the radius $R = D / 2 = 25 \, {\rm mm}$, and it is assumed that
the temperature $T$ inside the cylinder, for all times $t \geq 0$,
depends only on the radial distance $r \in [0, R]$ from the center of
the cylinder. The temperature distribution $T = T(r, t)$ inside the cylinder
is determined by the one-dimensional heat conduction equation (HCE),
written in polar coordinates without the angular component
\begin{equation}
	c \rho \, \frac{\partial T}{\partial t} =
		\frac{1}{r} \cdot \frac{\partial}{\partial r}
		\left(
		r \lambda \, \frac{\partial T}{\partial r}
		\right),
\label{Eq.2}
\end{equation}
where $c$, $\rho$, and $\lambda$ are thermal properties of the material:
\begin{itemize}
  \item[]
	$c$ = specific heat, ${\rm J / (kg \cdot K) }$,
  \item[]
	$\rho$ = density, ${\rm kg / m^3 }$, and
  \item[]
	$\lambda$ = thermal conductivity, ${\rm W / (m \cdot K) }$.
\end{itemize}
As will be shown in one of the sensitivity tests, these properties must
be taken as temperature dependent, i.e., as functions of $T$.

	Before each quenching test, the probe is initially heated to about
$850 \, {\rm {}^\circ C}$, and the initial condition $T(r, 0) = T_0(r)$ at
the start of the quenching process is known. Usually, it is assumed that
the probe is uniformly heated, so $T_0$ is constant.

	From the radial symmetry assumption, the boundary condition at
the center ($r = 0$) must be
\begin{equation}
          \frac{\partial T}{\partial r} = 0,
\label{Eq.3}
\end{equation}
and the HTC $\alpha$ at the surface depends only on time $t$. The boundary
condition Eq~1 at the surface ($r = R$) can then be written as
\begin{equation}
	q_s = - \lambda \, \frac{\partial T}{\partial r} =
		\alpha \left( T_s - T_x \right),
\label{Eq.4}
\end{equation}
where $T_s(t) = T(R, t)$ is the (calculated) surface temperature, and
$T_x$ is the measured quenchant temperature.

	The purpose of the probe is to provide further insight into the actual
quenching conditions at the surface. To this aim, the HTC is calculated from
local information, i.e., from the near-surface temperature $T_n$, by
the so-called solution extension. The remaining two measured curves are used
only to verify the computed temperature distribution, but not in the actual
HTC calculation.

	In this approach, the computed $\alpha$ is very sensitive to the local
surface conditions, and this may give somewhat poorer results far away from
the surface. On the other hand, it avoids the damping effect deeper in
the probe, it is quite fast in practice, and ideally suited for sensitivity
tests.

	The original temperatures are measured at discrete times, until some
final time $t_{\rm final}$ (${\rm s}$), when the probe is sufficiently cool.
Before the HTC calculation, the near-surface temperature is smoothed over
the whole time range $[ 0, t_{\rm final} ]$, and this smoothing is the only
regularization in the algorithm.

	The smoothed near-surface temperature curve, also denoted by $T_n$,
is the main input for the HTC calculation. The HCE (Eq~2) is written as
the heat conservation law for the cylinder
\begin{equation}
	\int_{0}^{R} \left(
		c \rho \, \frac{\partial T}{\partial t}
		\right) (r, t) \cdot r \d r
	= R \cdot \left(
		\lambda \, \frac{\partial T}{\partial r}
		\right) (R, t)
	= -R \cdot q_s(t),
	\quad t \geq 0.
\label{Eq.5}
\end{equation}
The numerical solution of Eq~5 is based on the finite volume method (FVM)
\cite{Mattheij-Rienstra-tenThijeBoonkkamp-2005}, because it is easy to
incorporate both boundary conditions Eqs~3 and 4 in terms of the heat flux
density, and, by doing so, avoid explicit numerical differentiation of
temperature at the surface in Eq~4.

	The whole computation is performed in four nested loops. From a global
perspective, the solution progresses in a series of discrete time levels $t_i$
(${\rm s}$), from the initial level $t_0 = 0$, until the final time is reached.
\begin{itemize}
  \item[1.]
	In the outermost loop, the algorithm advances by one time level
	per step, with the time step $\tau_i = t_i - t_{i - 1}$ (${\rm s}$).
	The solution at the new time level $t_i$ is computed from the solution
	at the previous level $t_{i - 1}$.
  \item[2.]
	At each time level $t_i$, the surface heat flux density $q_s(t_i)$
	is calculated iteratively, until the resulting temperature at
	the position $R - d_n$ is equal to the smoothed near-surface
	temperature $T_n(t_i)$. The numerical solution of this equation
	is done by the Brent--Dekker method \cite{Brent-1973}, and
	these iterations constitute the middle loop of the algorithm.
  \item[3.]
	For each value of the surface heat flux density $q_s(t_i)$, generated
	by the Brent--Dekker iterations, the remaining two loops compute
	the temperature distribution over the selected space
	discretization---finite volumes that determine the space grid, in
	the whole interval $[0, R]$, by solving the direct HCE problem,
	written as Eq~5 for all finite volumes. The first inner loop performs
	simple iterations to adjust all thermal properties to new
	temperatures at time $t_i$, until they are stabilized.
  \item[4.]
	In each of these iterations, thermal properties at all space-grid
	points are treated as known. The innermost loop then computes
	the approximate temperatures at all space-grid points, by solving
	a tridiagonal linear system of equations.
\end{itemize}

	Step 2 is the key step for solving the IHCP, as it computes
the unknown surface heat flux density $q_s(t_i)$ in the boundary condition.
Effectively, it performs a solution extension on the interval $[R - d_n, R]$
from the near-surface position to the surface, over the space gap of length $d_n$.
That determines the temperature distribution, including the surface
temperature $T_s(t_i)$. After that, the HTC $\alpha$, as a function of time,
is obtained directly from Eq~4, by using the input $T_x$ value.

	This method for the solution of the IHCP is especially adapted
for large probes with the reference thermocouple located very close to
the surface, so that the solution extension toward the surface is performed
over a small space gap with respect to the whole radius of the probe.
It is not suitable for small probes with a central thermocouple.

%
%
\section{General sensitivity concerns}
%
%
	Because the HTC calculation is an ill-posed problem, and high
sensitivity is expected, before doing any sensitivity tests, every effort has
to be made to recognize and possibly eliminate all issues that may influence
the validity of the results.

	An actual experiment by the probe is subject to many factors that
are beyond control, and may affect the results. As an example, the radial
symmetry assumption requires strictly vertical immersion of the probe,
and vertical agitation of the quenchant. Both assumptions are often
violated in practice, resulting in an asymmetrical temperature distribution.
For such reasons, sensitivity tests must be conducted in a carefully
controlled environment, that is not affected by experimental errors, and they
cannot be done by comparing the computed results with measured temperatures.
The only reasonable way to do them is by numerical simulation, i.e., by using
the same method that is used for the HTC calculation.

	The goal of the sensitivity analysis is to determine just how sensitive
are various output results with respect to reasonably small perturbations of
the input data. These results are used to judge the reliability and numerical
stability of the method, and how accurate is the computation of output
from input.

	However, this should not be confused with the true accuracy of
the results. For example, the computed HTC may be nowhere near the true value,
which is quite hard to find, anyway. That has to be verified
independently---by practical experiments and other computational methods.
Once it has been verified that the results are reasonably accurate,
then the sensitivity results for the calculation procedure can be extended
into some conclusions about the sensitivity of the HTC itself.
In that respect, a comparison of the other two measured temperatures
with the computed results \cite{Liscic-Singer-2013,Liscic-Singer-2014} confirms
that the results are satisfactory, within the limits of the simplified 1D
model used for calculation.

%
%
\subsection{General sensitivity of the method}
%
%
	To obtain good results, before the test, all the data has to be
prepared carefully, to avoid any potential numerical instabilities in
the method, that may be caused by inappropriate data, and the method has to
be guarded against such instabilities, as much as possible.

	Most of the numerical methods for solving the HCE, including the FVM
and the finite difference methods, are very sensitive to the lack of
smoothness in the initial and boundary conditions, and in the coefficients
of the equation, i.e., in the thermal properties of the material. The cause
of that lies in the assumed smoothness of the solution (temperature) that is
required to ensure the validity of the equation that is being solved---in
this case, Eq~2 or Eq~5. The standard smoothness requirement is that
the temperature must be continuously differentiable in time (the so-called
$C^1$ smoothness). It should be noted that this type of sensitivity has
nothing to do with the ill-posedness of the HTC calculation.

%
%
\subsection{Smoothness of input data}
%
%
	The method used for the HTC calculation is extremely sensitive to
the smoothness of the near-surface temperature curve $T_n(t)$, for
several additional reasons, related to the ill-posedness of the problem.
This curve is the only input used in the essential part of the computation
to determine the surface heat flux density $q_s(t_i)$, and $T_x$ is used
only to calculate $\alpha(t_i)$ from $q_s(t_i)$.
The surface conditions are estimated from local information, and $T_n(t_i)$
plays the role of an ``internal'' reference condition for the solution
extension. As the gap is very small, every small wiggle in the input
$T_n(t)$ curve is reflected and magnified in the computed $\alpha(t)$ curve.
Because there is no other (explicit or implicit) regularization involved
in the solution of the IHCP, the near-surface temperature curve $T_n(t)$
must be sufficiently smooth---at least continuously differentiable
($C^1$ smoothness). Therefore, the smoothing of $T_n$ is an essential
part of the data preparation.

	It is interesting that the results, including the HTC, are not very
sensitive to the choice of the smoothing method, as long as the input curve
is sufficiently smooth, and the errors in the approximation of the original
measured data are sufficiently small.
For example, the reference cooling curves in the next section are prepared
by using the cubic spline least-squares approximation \cite{Dierckx-1993},
which is twice continuously differentiable ($C^2$ smoothness), while
the software that comes with the probe uses exponential $C^1$ splines
\cite{Liscic-Singer-2013,Liscic-Singer-2014}.
The differences in the HTC are almost negligible, and strictly confined to
the high temperature gradients portion of the quenching process.

	On the other hand, all local smoothing methods, such as moving
averages and local polynomial least-squares approximations, also known as
the Savitzky-Golay filters \cite{Savitzky-Golay-1964}, just damp the noise
in the data, and the resulting curve may turn out to be insufficiently smooth.

	At the beginning of the calculation---at the first time level, it is
important to ensure a smooth transition from the initial condition to
the boundary or the reference condition, that is used later on. Consequently,
to avoid any jumps at the start, the actual value of the initial temperature
$T_0$ is taken from the smoothed temperature curve as $T_0 = T_n(0)$.
Because the initial temperature is assumed to be constant over the whole
interval $[0, R]$, the heat flux density $q$ (${\rm W / m^2}$) between all
control volumes, and the initial surface heat flux density $q_s(0)$ must all
be equal to zero.
Moreover, the same must hold for the first derivative of the temperature,
$T'_n(0) = 0$, and this constraint is added in the initial smoothing and
approximation of the near-surface temperature curve.

	The experience shows that even a small violation of these conditions
causes severe oscillations of the computed solution near the surface for
the first few time steps. This may be seen as the numerical instability of
the method, but it is just an expected reaction to non-smooth conditions,
and can be avoided by proper data preparation.

	A similar thing can happen at any time, particularly when intensive
cooling begins. All sudden changes in temperature must be $C^1$-smooth,
at least on the time scale involved in the numerical solution.

%
%
\subsection{Time advancement and numerical stability}
%
%
	When such oscillations in the solution appear, what happens
afterwards, depends on the choice of the time advancement method in
the outermost loop of the algorithm. There are three standard choices for
the numerical solution of the HCE
\cite{Mattheij-Rienstra-tenThijeBoonkkamp-2005}, and they will be described
with their usual names, and in terms of numerical time-integration rules
(which are used in the FVM implementation).
\begin{itemize}
  \item[1.]
	Explicit method (first-point rule for time integration) is stable,
	i.e., damps these oscillations, only when extremely small time steps
	are used, and the calculation becomes prohibitively slow.
  \item[2.]
	Implicit method (last-point rule for time integration) is
	unconditionally stable, so large time steps can be used. Any
	oscillations are relatively quickly damped in a few time steps,
	but the damage is done in all computed values. Typically, the surface
	temperature $T_s(t)$ is not monotone, and the jumps are amplified in
	$q_s(t)$ and $\alpha(t)$ curves. An additional smoothing may be
	required to remove these artificial oscillations.
  \item[3.]
	Crank--Nicolson method (trapezoidal rule for time integration) is,
	nominally, more accurate than the previous methods. However, it turned
	out to be very sensitive and unstable---the oscillations always appear
	and quickly blow up, causing numerical overflow exception.
\end{itemize}
The implicit method is used in all HTC calculations. Eventual numerical
oscillations can be avoided by appropriate smoothing and handling of data.
In all the tests reported here, such instabilities have never occured.

%
%
\subsection{Smoothness of thermal properties}
%
%
	The values of thermal properties, that appear as coefficients in
the HCE, are known only for a small number of temperatures. Before use, these
tables have to be converted into functions of temperature over the whole
temperature range. Two such functions are needed, one for the so-called
volumetric specific heat $\gamma = c \rho$ (${\rm J / (m^3 \cdot K) }$),
and one for the thermal conductivity $\lambda$.
Based on the data from \cite{Inconel600-2008,Clark-Tye-2003} for Inconel 600,
these functions have been computed by Akima's piecewise cubic $C^1$
interpolation \cite{Akima-1970}.

	The same can be done by any other form of interpolation (or even
approximation) that gives the global $C^1$ smoothness, such as cubic spline
interpolation \cite{deBoor-2001}, but not with piecewise linear interpolation.

	Even though it is the most commonly used method for converting tables
into functions, the piecewise linear interpolation should be avoided in
the HTC calculation, not only for thermal properties, but also for
temperatures in time or space. The resulting curve is continuous, but its
derivative has jumps at the interpolation nodes, so it is not $C^1$-smooth.
As a result, such interpolation of thermal properties may drastically
increase the number of Brent--Dekker and simple iterations required in
steps~2 and~3 of the algorithm.

%
%
\section{Design and scope of numerical experiments}
%
%
	Numerical experiments for sensitivity tests consist of computing
and comparing relevant output results for different values of various
input parameters of the problem.

	Strictly speaking, the HTC calculation has only two inputs:
the smoothed near-surface cooling curve $T_n(t)$, and the (constant)
quenchant temperature $T_x$. In practice, everything else is fixed, as
determined by the design of the probe.
The main advantage of numerical experiments is that they can go beyond that.
Anything that can be considered as an input, or as a parameter of the problem,
can be changed and varied, including all numerical parameters in the method.

	In this approach to the HTC calculation, the input cooling curve
characterizes the whole quenching process in a particular quenchant, and
it is obvious that the surface heat flux density and the HTC critically
depend on it. But, this dependence is not the target of the tests.
Taking different input cooling curves is equivalent to changing quenchants
and quenching conditions, and that must be based on the actual data, not
on numerical simulation.

	Instead of playing with fictive curves representing fictive
quenchants, to avoid any misinterpretation of the results, only two cooling
curves with fixed corresponding quenchant temperatures are used in all tests.
They have been chosen to represent two different liquid quenchants that lie
at the opposite ends of quenching intensity:
\begin{itemize}
  \item
	mineral oil, non-accelerated and without agitation, with the peak
	HTC value of about $2{,}800 \ {\rm W / (m^2 \cdot K) }$, and
  \item
	water, with the peak HTC value of almost
	$10{,}000 \ {\rm W / (m^2 \cdot K) }$.
\end{itemize}
Most other liquid quenchants are somewhere in between these two.
A smoothed reference cooling curve has been prepared for each quenchant,
based on several measurements by the probe, and then smoothed with all
the necessary constraints, as described in the previous section. This means
that only two different inputs (in the strict sense) are used for all tests.

	With these two inputs, that represent two actual quenchants, the main
objective of numerical tests is to find the sensitivity of output with
respect to all other parameters of the problem, that cannot be changed easily
in practice.

	The following five groups of experiments have been conducted by
performing a selected set of variations of the corresponding parameter:
\begin{itemize}
  \item[1.]
	Thermal properties of the material, used in the HCE to calculate
	the results.

  \item[2.]
	The depth $d_n$ of the near-surface thermocouple, without changing
	the reference temperatures $T_n$ and $T_x$, as if the same
	temperatures $T_n(t)$ are measured at a different position.

  \item[3.]
	The diameter $D$ of the probe, with the same depth $d_n$.

  \item[4.]
	The space step $h$, to test the sensitivity of the calculation
	(the innermost loop of the algorithm).

  \item[5.]
	The time step $\tau$ between successive time levels, to test
	the sensitivity of the calculation (the outermost loop of
	the algorithm).
\end{itemize}
The underlying motivation for each group of tests will be discussed along
with the results in the next section.

	The usual output of the calculation consists of HTC and surface
heat flux density curves, together with a number of calculated temperature
curves at various points on the radius of the probe. For the purpose of
sensitivity analysis, just four of these curves will be considered,
and all of them as a function of time $t$:
%
%
\begin{itemize}
  \item calculated core temperature (for $r = 0$), denoted by $T_c$
	(${\rm {}^\circ C}$),
  \item calculated surface temperature (for $r = R$), denoted by $T_s$
	(${\rm {}^\circ C}$),
  \item calculated HTC $\alpha$,
  \item calculated surface heat flux density $q_s$.
\end{itemize}
The core temperature $T_c$ is taken as a representative of what happens far
away from the surface, while the last three curves represent the behavior
at the surface.
%
%
%

	There are several good reasons why the HTC is considered only
as a function of time. In gas quenching, which is controllable in time,
this the natural state of affairs. On the other hand, in liquid quenching,
the HTC is often taken as a function of the surface temperature $T_s$.
But, from the computational point of view, the HTC is always computed
first as a function of time, and then converted into a function of $T_s$,
provided that the surface temperature is monotone decreasing in time.
Unfortunately, for some polymer solutions, this is not always the case
\cite{Liscic-Singer-2013,Liscic-Singer-2014}, and requires additional
efforts to get the HTC as a function of $T_s$. Finally, in sensitivity
tests, it is much easier to compare the differences in computed HTCs
on the fixed time-scale, when only one thing is changed, not two.

%
%
	The sensitivity is measured in the absolute sense---as the difference
between the computed results. To simplify the whole procedure, only one set
of output results, computed with standard values of all parameters, will be
taken as the reference set for each quenchant (cooling curve). All other
results for perturbed data will be compared with these reference results,
and only the difference between the reference curve and the ``perturbed''
curve will be shown.

	The results will be presented in a graphical format, in hope that
a picture tells a thousand words. The essential information is contained in
the scale on the $y$-axis (that gives the magnitude of the difference, i.e.,
the result of the perturbation), and in the shape of the difference curve.

%
%
\subsection{Reference data for oil}
%
%
	The first reference set represents the probe quenched in a
low-viscosity plain (non-accelerated) quenching oil, without agitation.
The input near-surface temperature curve $T_n(t)$ and the quenchant
temperature $T_x  = 66 \, {\rm {}^\circ C}$ are shown in left part of
Fig.~\ref{Oil_Ref_Temp} for the whole quenching period of $900$ seconds.
The calculated temperatures $T_s(t)$ at the surface, and $T_c(t)$ at the core,
are also shown in the same figure. These two temperatures are taken as
the reference curves in all subsequent tests for oil.

	The quenching process is quite slow, and intensive cooling starts
relatively late, after about $36$ seconds. The right part of
Fig.~\ref{Oil_Ref_Temp} shows all the temperatures during the most intensive
part of the quenching period, in the fine time-scale. The same style of dual
time-scale presentation will be used through, where appropriate.

\begin{center}
	\epsfbox{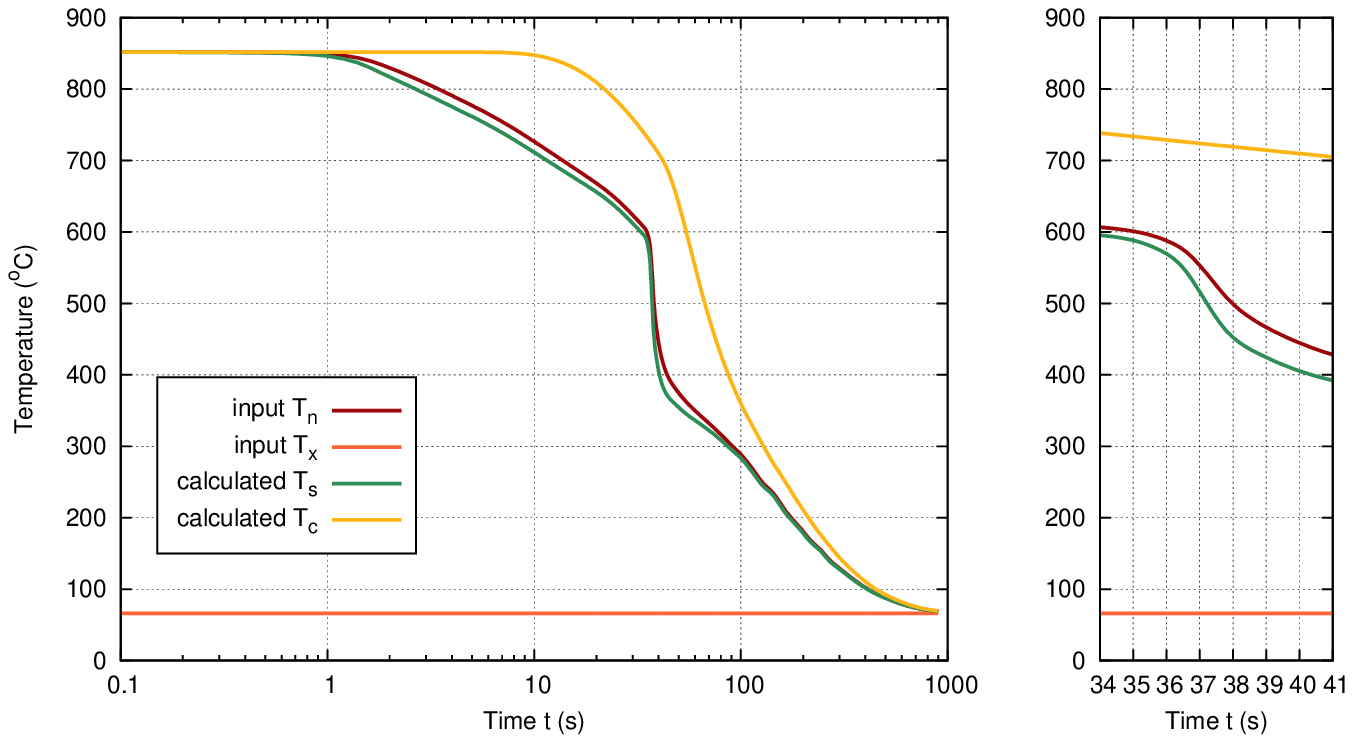}
\end{center}%
\Fig{Oil_Ref_Temp}%

\vspace*{-3pt}
\noindent
\hbox{Fig.~\ref{Oil_Ref_Temp} \ }%
	Oil (reference): input near-surface and quenchant temperatures,
	and calculated surface and core temperature curves.
\vspace*{12pt}
%

	The calculated HTC and the surface heat flux density $q_s$ are shown
in Figs.~\ref{Oil_Ref_Alpha} and~\ref{Oil_Ref_Flux}, and they are taken as
the corresponding reference curves in all tests for oil.

\begin{center}
	\epsfbox{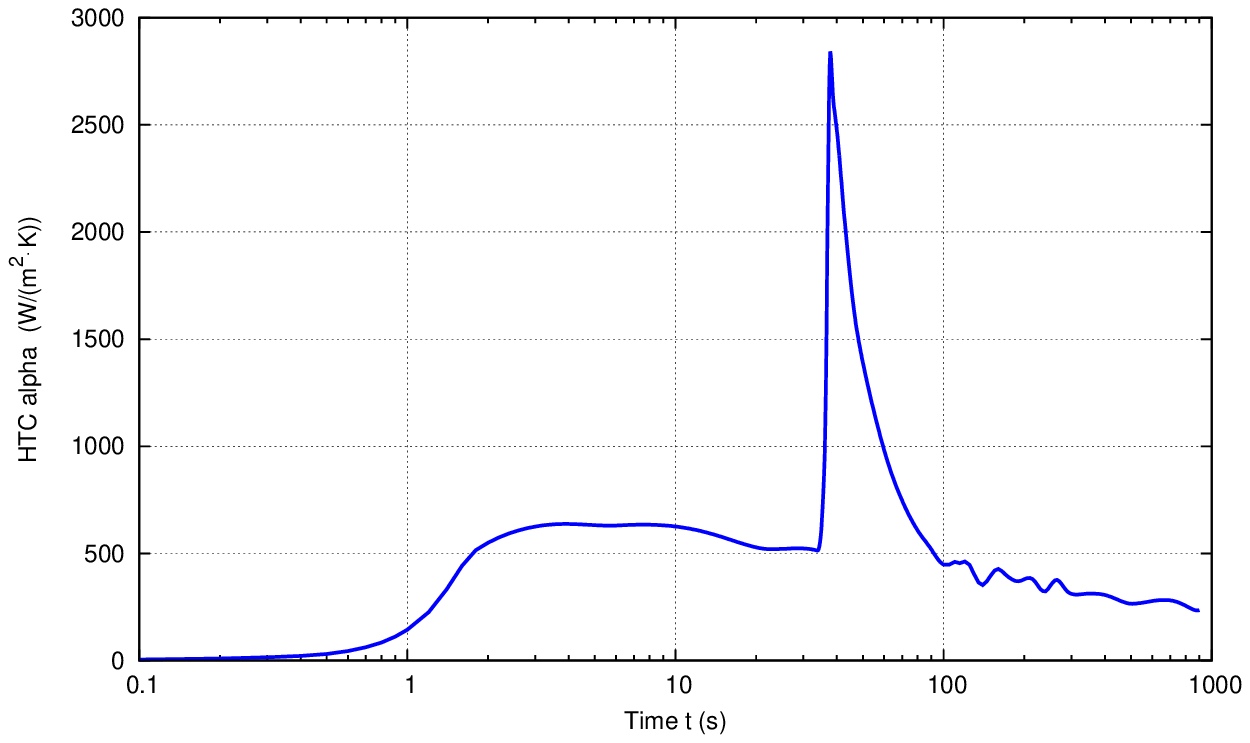}
\end{center}%
\Fig{Oil_Ref_Alpha}%

\vspace*{-3pt}
\noindent
\hbox{Fig.~\ref{Oil_Ref_Alpha} \ }%
	Oil (reference): calculated HTC as a function of time.
\vspace*{12pt}
%

\begin{center}
	\epsfbox{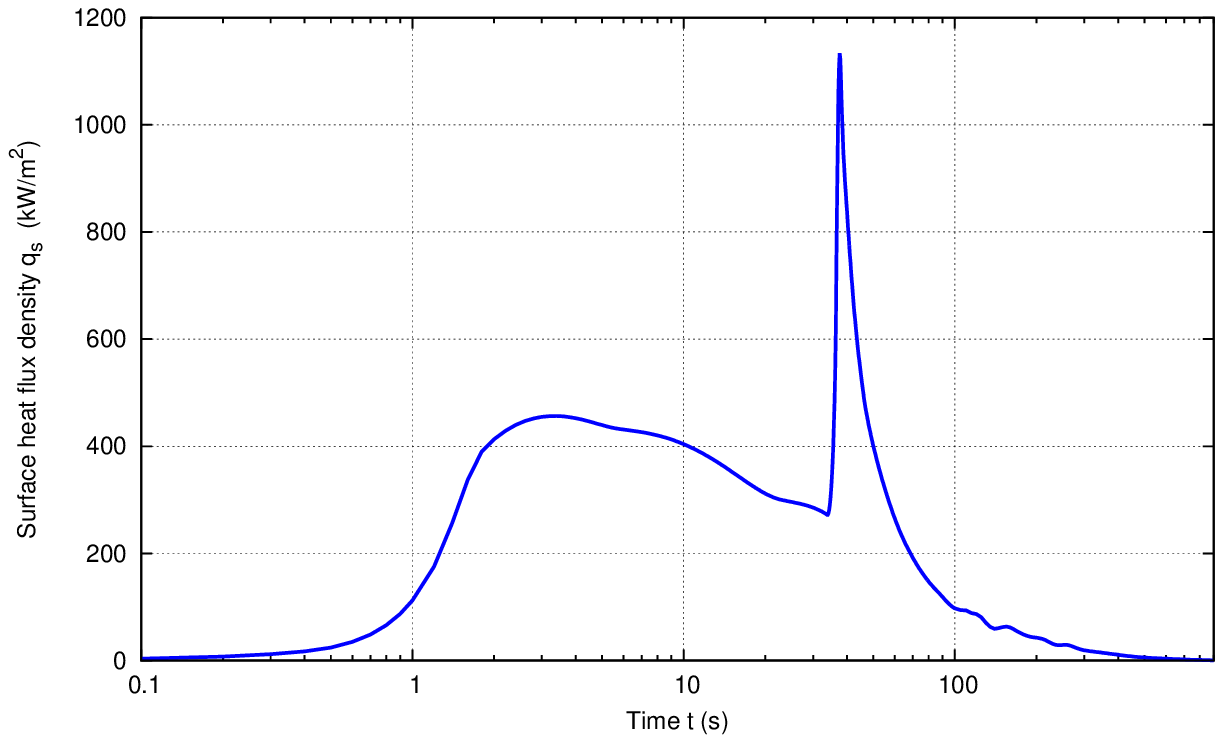}
\end{center}%
\Fig{Oil_Ref_Flux}%

\vspace*{-3pt}
\noindent
\hbox{Fig.~\ref{Oil_Ref_Flux} \ }%
	Oil (reference): calculated surface heat flux density as a function
	of time.
\vspace*{12pt}
%

%
%
\subsection{Reference data for water}
%
%
	The second reference set represents the probe quenched in water
at the temperature $T_x = 38 \, {\rm {}^\circ C}$.
The input near-surface temperature curve $T_n(t)$ and $T_x$ are shown in
the left part of Fig.~\ref{Wat_Ref_Temp}, together with the calculated
surface and core temperatures, $T_s(t)$ and $T_c(t)$ (the reference curves
for water), for the whole quenching period of $300 \, {\rm s}$.

	The quenching in water is much quicker, and intensive cooling starts
quite soon, after only $10$ seconds. Again, the right part of
Fig.~\ref{Wat_Ref_Temp} shows all the temperatures during the most
intensive part of the quenching period, in the fine time-scale.

\begin{center}
	\epsfbox{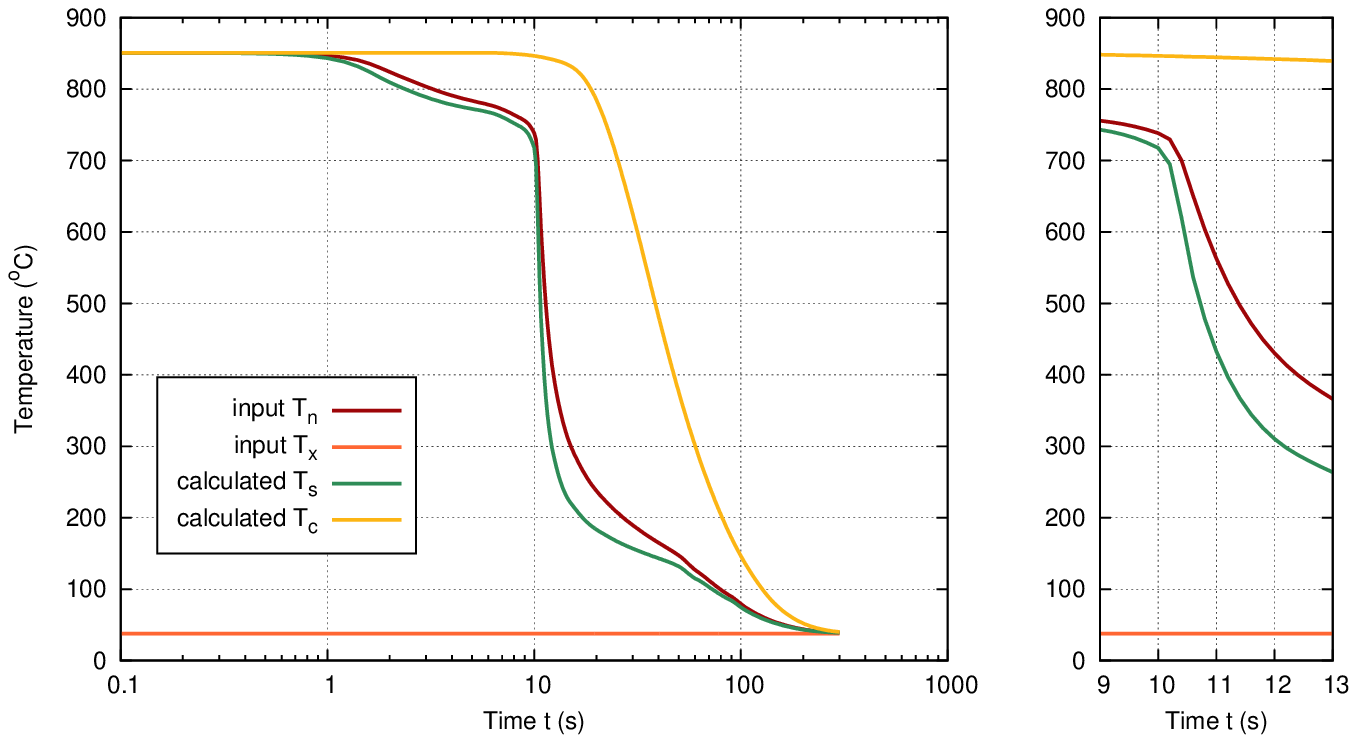}
\end{center}%
\Fig{Wat_Ref_Temp}%

\vspace*{-3pt}
\noindent
\hbox{Fig.~\ref{Wat_Ref_Temp} \ }%
	Water (reference): input near-surface and quenchant temperatures,
	and calculated surface and core temperature curves.
\vspace*{12pt}
%

	The calculated HTC and the surface heat flux density $q_s$ are shown
in Figs.~\ref{Wat_Ref_Alpha} and~\ref{Wat_Ref_Flux}, and they are taken as
the corresponding reference curves in all tests for water.

\begin{center}
	\epsfbox{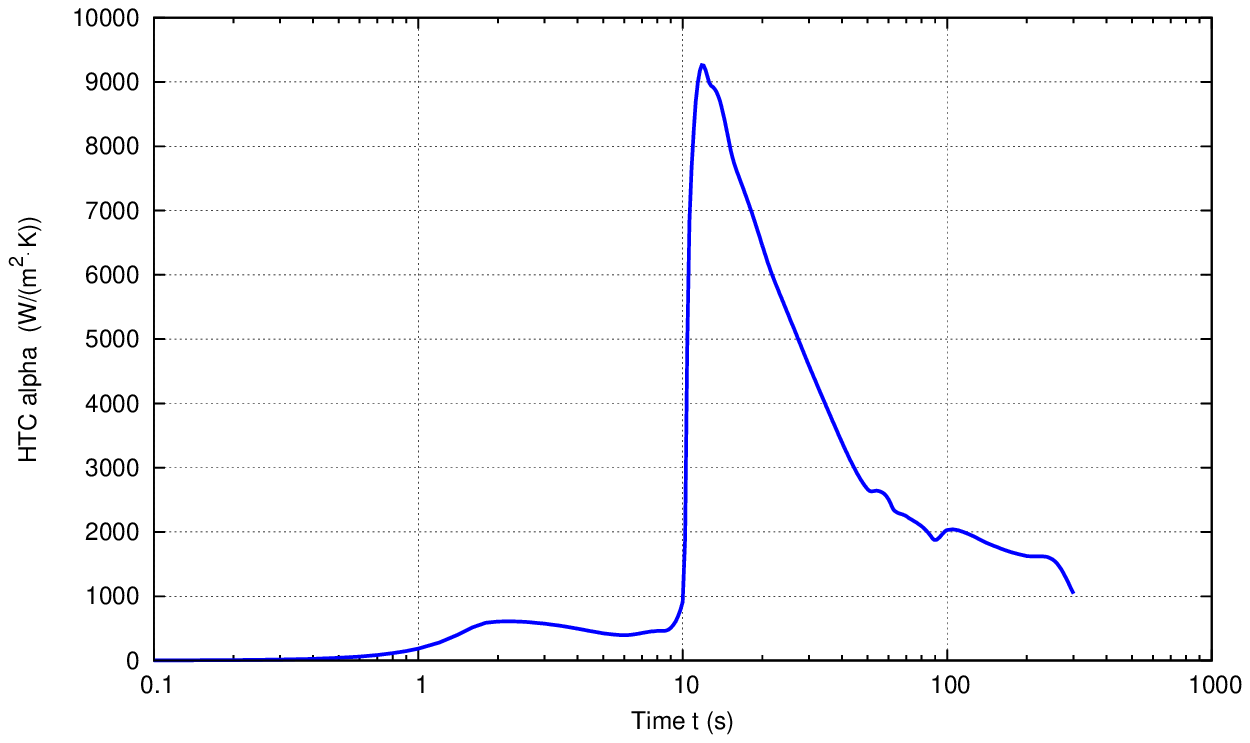}
\end{center}%
\Fig{Wat_Ref_Alpha}%

\vspace*{-3pt}
\noindent
\hbox{Fig.~\ref{Wat_Ref_Alpha} \ }%
	Water (reference): calculated HTC as a function of time.
\vspace*{12pt}
%

\begin{center}
	\epsfbox{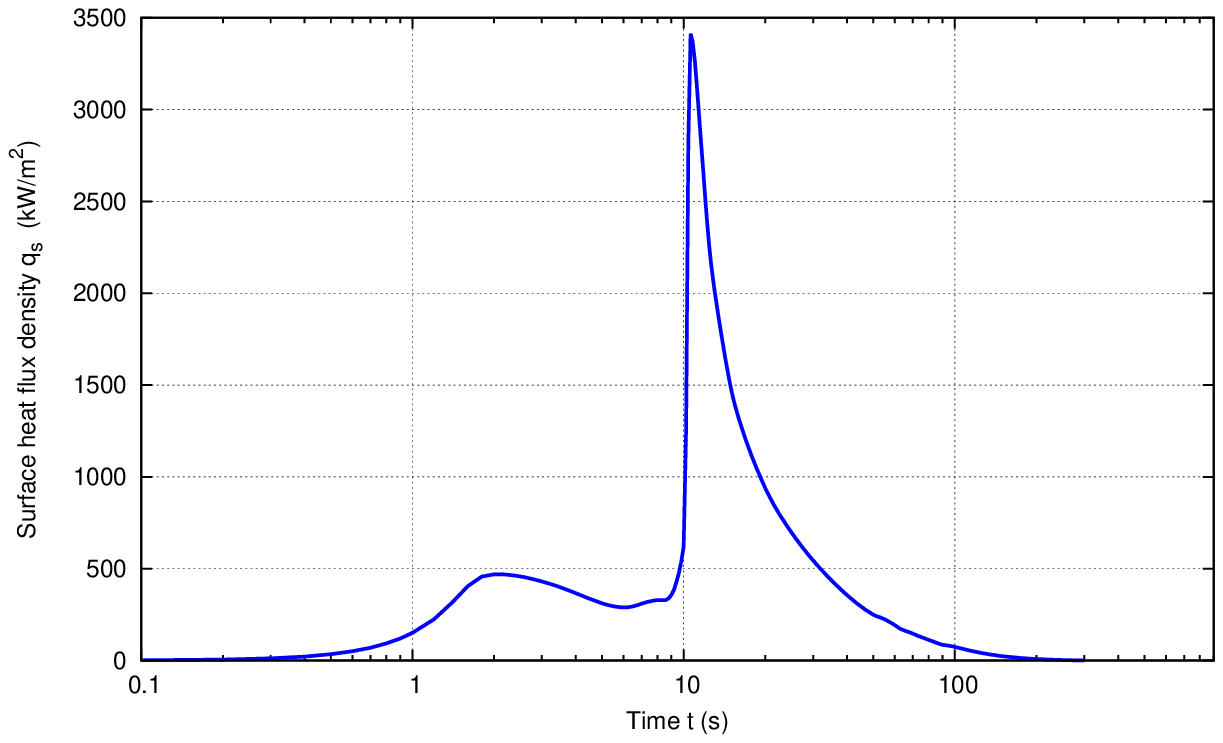}
\end{center}%
\Fig{Wat_Ref_Flux}%

\vspace*{-3pt}
\noindent
\hbox{Fig.~\ref{Wat_Ref_Flux} \ }%
	Water (reference): calculated surface heat flux density as a function
	of time.
\vspace*{12pt}
%

	In both HTC curves, shown in Figs.~\ref{Oil_Ref_Alpha}
and~\ref{Wat_Ref_Alpha}, small waves or wiggles can be seen relatively
near the end of the quenching process (in the logarithmic time-scale),
starting around $130$ seconds for oil, and around $50$ seconds for water.
They are caused by the actual phenomenon at the surface, when the quenchant
layer near the surface becomes sufficiently cool to stop boiling, and only
sporadic vapor bubbles are formed, before it turns into the entirely liquid
state.

%
%
\section{Results of sensitivity tests}
%
%
	This sections contains the computed results for all five groups of
sensitivity tests. The results for each group are presented in a separate
subsection.

%
%
\subsection{Thermal properties of the material}
%
%
	All thermal properties $c$, $\rho$, and $\lambda$ of the material
in Eqs~2 and 5, depend on the current temperature at a particular point
in the probe. During the numerical solution of the HCE, they have to be
adjusted repeatedly, to reflect the correct thermal behavior. Therefore,
the HTC calculation contains an additional loop (step~3 of the algorithm)
to perform this adjustment by simple iterations. As a consequence, the whole
computation is slowed down significantly.

	In order to speed up the calculation, it is quite popular in practice
to take constant values of thermal properties at some fixed temperature.
One of the objectives of this test is to see how that affects the solution.
The output curves are computed by taking all thermal properties for
Inconel 600 as constant values, first at the room temperature of
$30 \, {\rm {}^\circ C}$, and then at $400 \, {\rm {}^\circ C}$, which is much
closer to the actual temperature of intensive cooling. This test, of course,
is impossible to do in practice, but it is essential for the HTC calculation
procedure.

	On the other hand, for practical applications, it is far more
interesting to see what happens when the material of the probe is changed
from one steel to another, which is quite expensive to do in real life.
So, in the third test, the probe is considered to be made of AISI 304
stainless steel, instead of Inconel 600. Because AISI 304 is composed
mostly of iron, the actual thermal properties are somewhat different
at the same temperatures \cite{Totten-Dossett-Kobasko-2013}, but both steels
exhibit quite similar behavior when quenched.

	The results of these three tests for quenching in oil are shown in
Figs.~\ref{Oil_Mat_Tc}--\ref{Oil_Mat_Flux}, and in
Figs.~\ref{Wat_Mat_Tc}--\ref{Wat_Mat_Flux}, for quenching in water.

\begin{center}
	\epsfbox{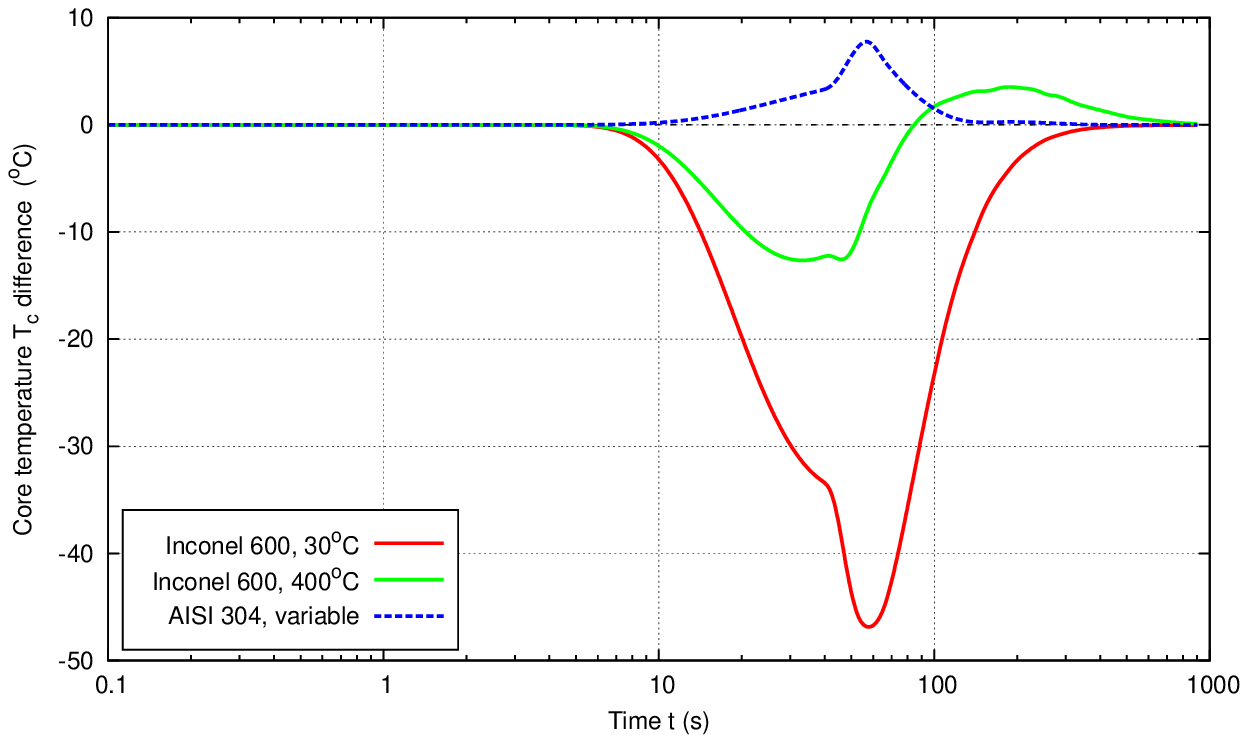}
\end{center}%
\Fig{Oil_Mat_Tc}%

\vspace*{-3pt}
\noindent
\hbox{Fig.~\ref{Oil_Mat_Tc} \ }%
	Oil (thermal properties variation):
	core temperature $T_c$ differences.
\vspace*{12pt}
%

\begin{center}
	\epsfbox{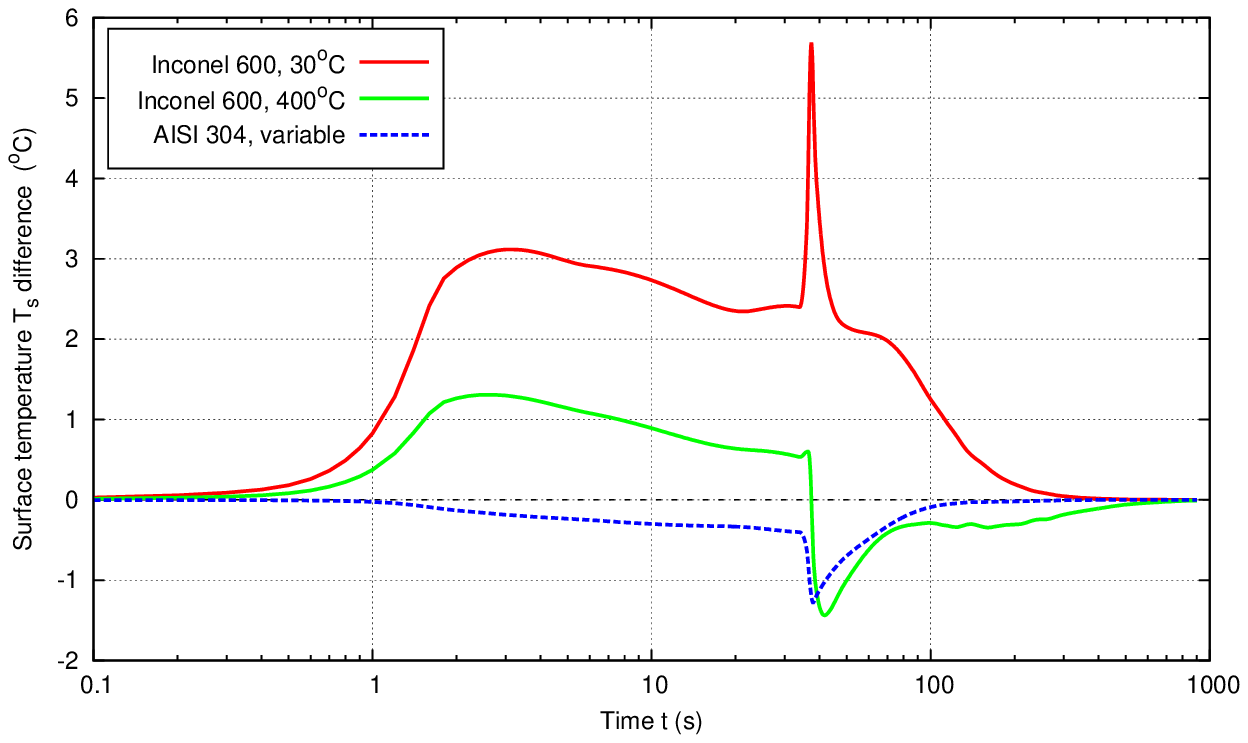}
\end{center}%
\Fig{Oil_Mat_Ts}%

\vspace*{-3pt}
\noindent
\hbox{Fig.~\ref{Oil_Mat_Ts} \ }%
	Oil (thermal properties variation):
	surface temperature $T_s$ differences.
\vspace*{12pt}
%

\begin{center}
	\epsfbox{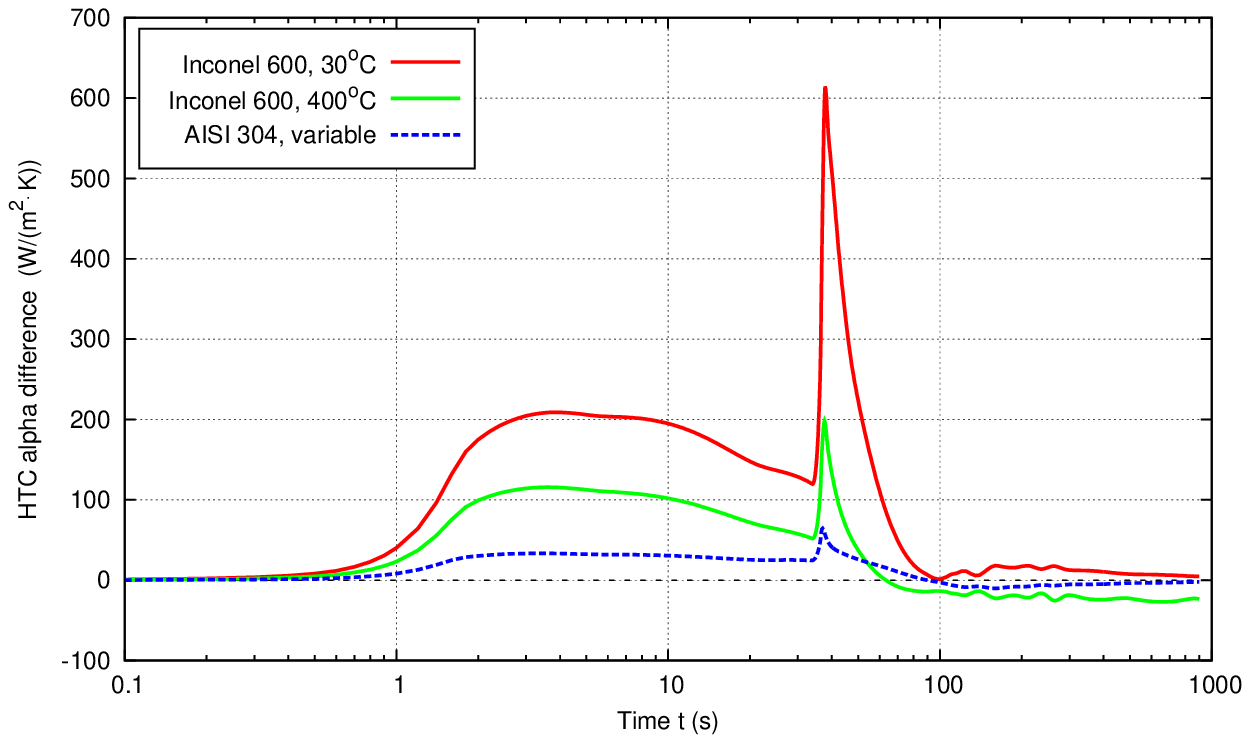}
\end{center}%
\Fig{Oil_Mat_Alpha}%

\vspace*{-3pt}
\noindent
\hbox{Fig.~\ref{Oil_Mat_Alpha} \ }%
	Oil (thermal properties variation):
	HTC $\alpha$ differences.
\vspace*{12pt}
%

\begin{center}
	\epsfbox{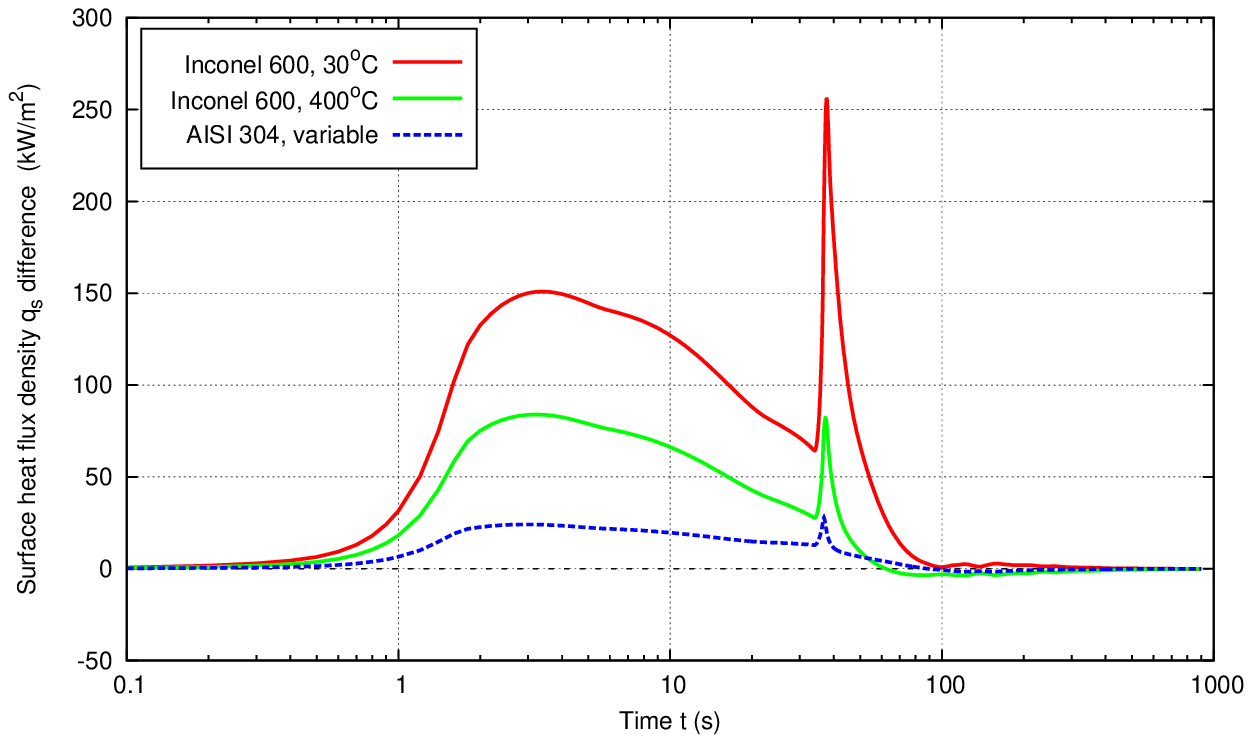}
\end{center}%
\Fig{Oil_Mat_Flux}%

\vspace*{-3pt}
\noindent
\hbox{Fig.~\ref{Oil_Mat_Flux} \ }%
	Oil (thermal properties variation):
	surface heat flux density $q_s$ differences.
\vspace*{12pt}
%

\begin{center}
	\epsfbox{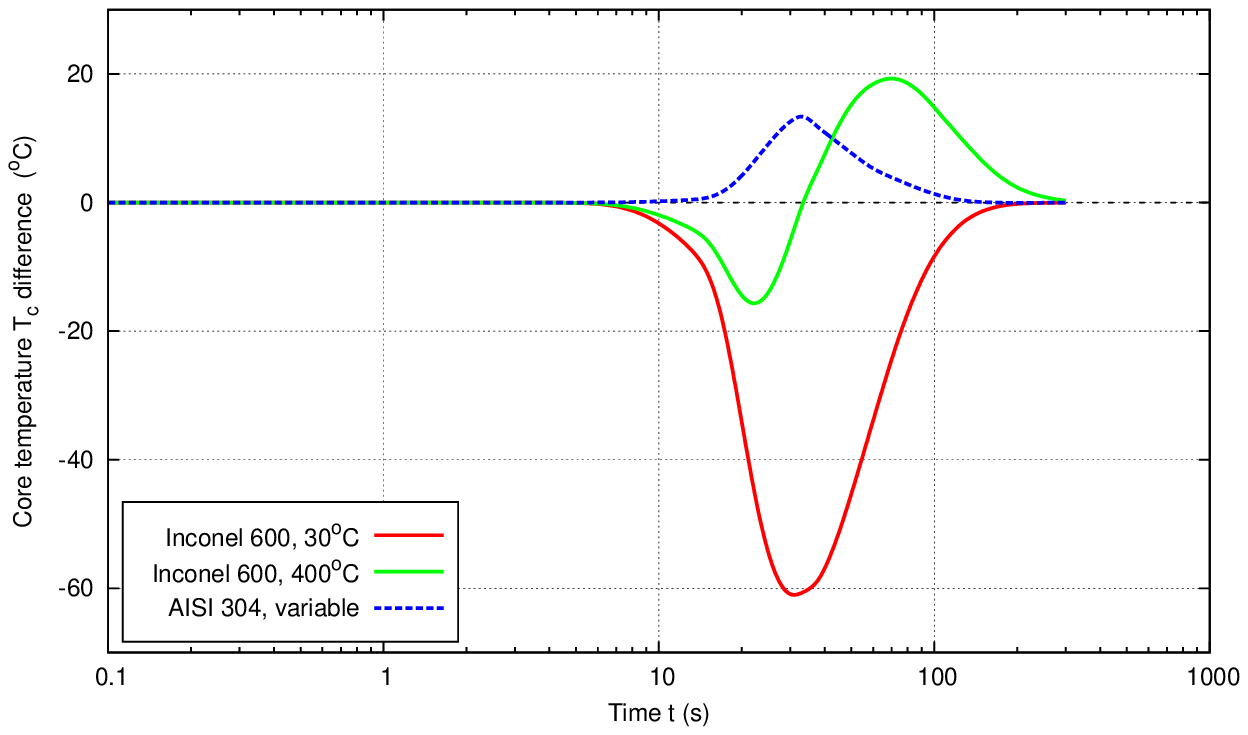}
\end{center}%
\Fig{Wat_Mat_Tc}%

\vspace*{-3pt}
\noindent
\hbox{Fig.~\ref{Wat_Mat_Tc} \ }%
	Water (thermal properties variation):
	core temperature $T_c$ differences.
\vspace*{12pt}
%

\begin{center}
	\epsfbox{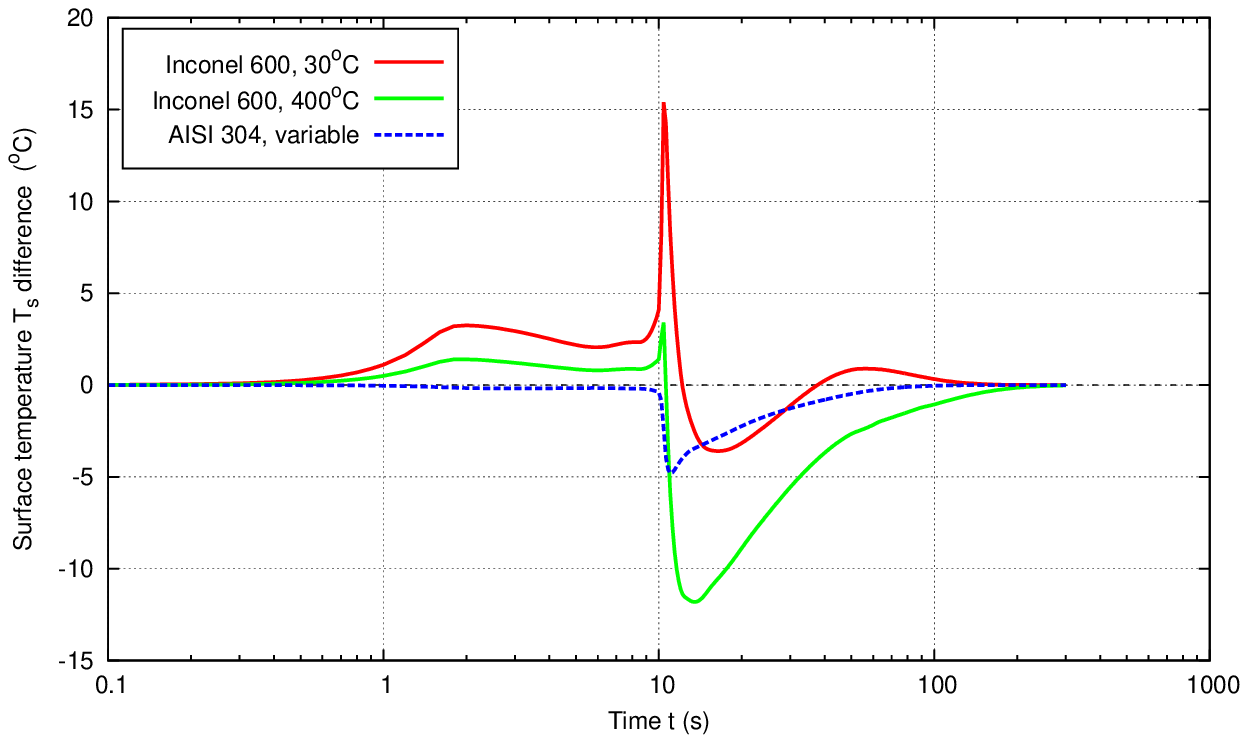}
\end{center}%
\Fig{Wat_Mat_Ts}%

\vspace*{-3pt}
\noindent
\hbox{Fig.~\ref{Wat_Mat_Ts} \ }%
	Water (thermal properties variation):
	surface temperature $T_s$ differences.
\vspace*{12pt}
%

\begin{center}
	\epsfbox{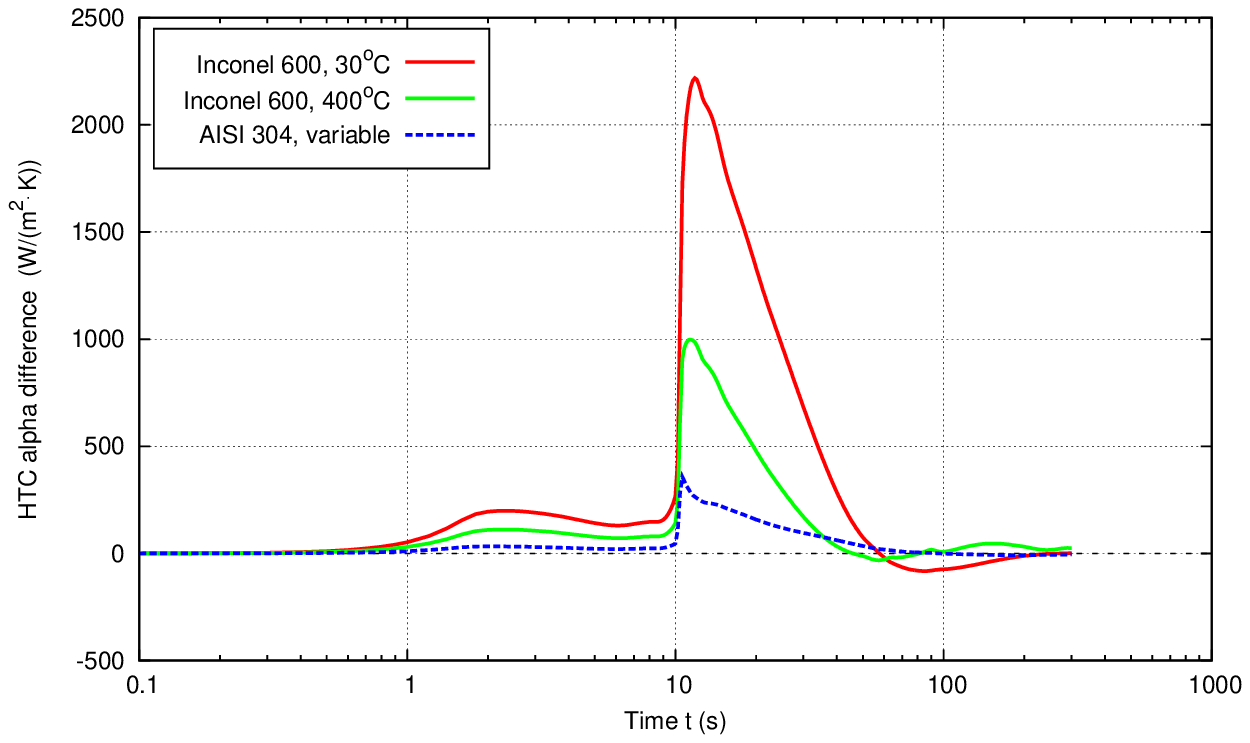}
\end{center}%
\Fig{Wat_Mat_Alpha}%

\vspace*{-3pt}
\noindent
\hbox{Fig.~\ref{Wat_Mat_Alpha} \ }%
	Water (thermal properties variation):
	HTC $\alpha$ differences.
\vspace*{12pt}
%

\begin{center}
	\epsfbox{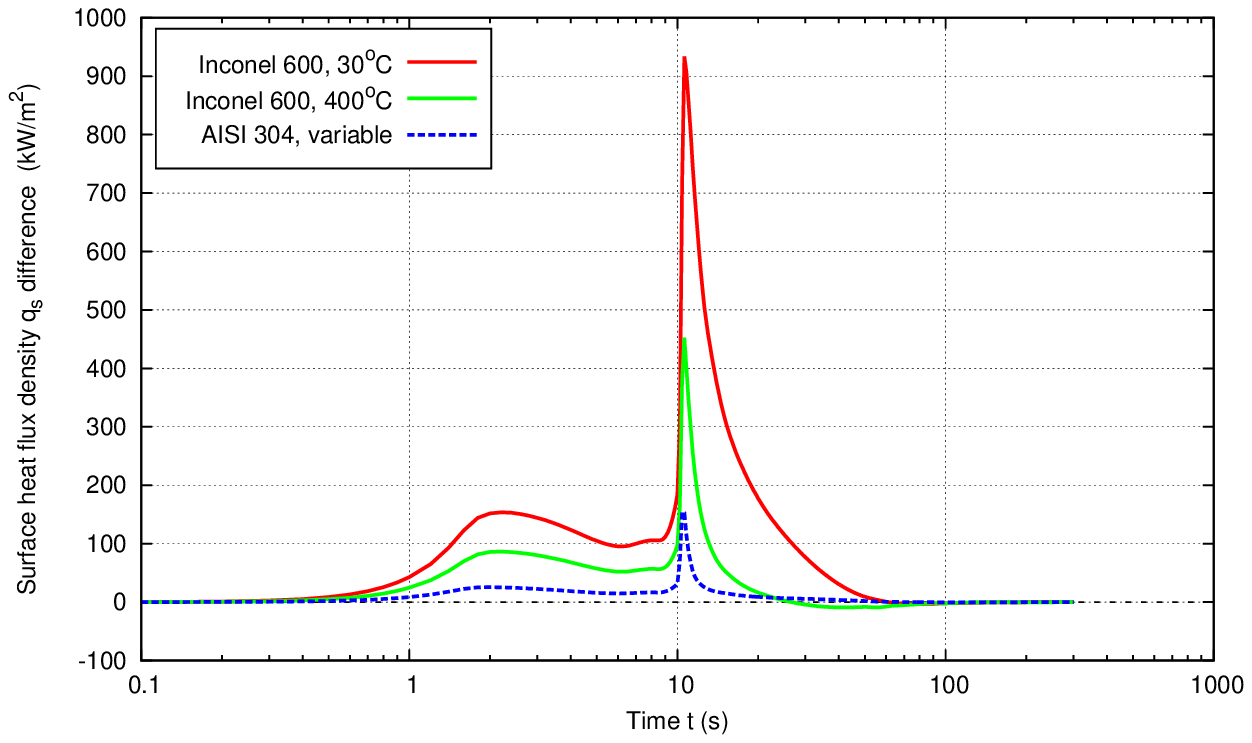}
\end{center}%
\Fig{Wat_Mat_Flux}%

\vspace*{-3pt}
\noindent
\hbox{Fig.~\ref{Wat_Mat_Flux} \ }%
	Oil (thermal properties variation):
	surface heat flux density $q_s$ differences.
\vspace*{12pt}
%

	In all output curves, the highest differences occur during the most
intensive cooling period, and this behavior is typical for all subsequent
tests, as well. Constant thermal properties, especially at the room
temperature, result in very high differences. For example, the peak HTC
values, both in oil and water, are underestimated by about~$25 \%$.
The test definitely shows that all thermal properties must be considered
as temperature-dependent in quenching simulations.

	More interestingly, the change of the material by another steel that
is known to have similar quenching characteristics, results in relatively
small differences in all output curves, i.e., the simulation confirms that
they are similar (which verifies the simulation itself).
This comment is valid only for similar steels, and a completely different
steel would result in much higher differences, but the point is that they
can be predicted by simulation.

%
%
\subsection{Near-surface thermocouple depth}
%
%
	The reference curves are computed from the near-surface temperature
$T_n$ measured at the depth $d_n = 1 \, {\rm mm}$ below the surface.
When the probe is manufactured, it is quite hard to achieve the exact drilling
and positioning of the near-surface thermocouple, and some small differences
in the depth can be expected in practice. Moreover, the diameter of
the thermocouple is about $1 \, {\rm mm}$, and it is very hard to tell what
is the actual depth that corresponds to the measured temperature $T_n$.

	The goal of this test is to study the effects of the depth variation,
and six different depths $d_n$ are used, with the step of $0.1 \, {\rm mm}$,
three of them above, and three below the standard position. The same
near-surface temperature curve $T_n(t)$ is used, as in the reference results,
even though it is very unlikely that the same temperature curve would
be measured at different positions in the same quenching conditions.
This is the only sensible thing to do, without actually doing the experiments
with such probes. Besides, the aim is to analyze the sensitivity with respect
to the depth, so only $d_n$ should be varied.

	The results of all six tests for quenching in oil are shown in
Figs.~\ref{Oil_Tnp_Tc}--\ref{Oil_Tnp_Flux}, and the results in water are
shown in Figs.~\ref{Wat_Tnp_Tc}--\ref{Wat_Tnp_Flux}.

\begin{center}
	\epsfbox{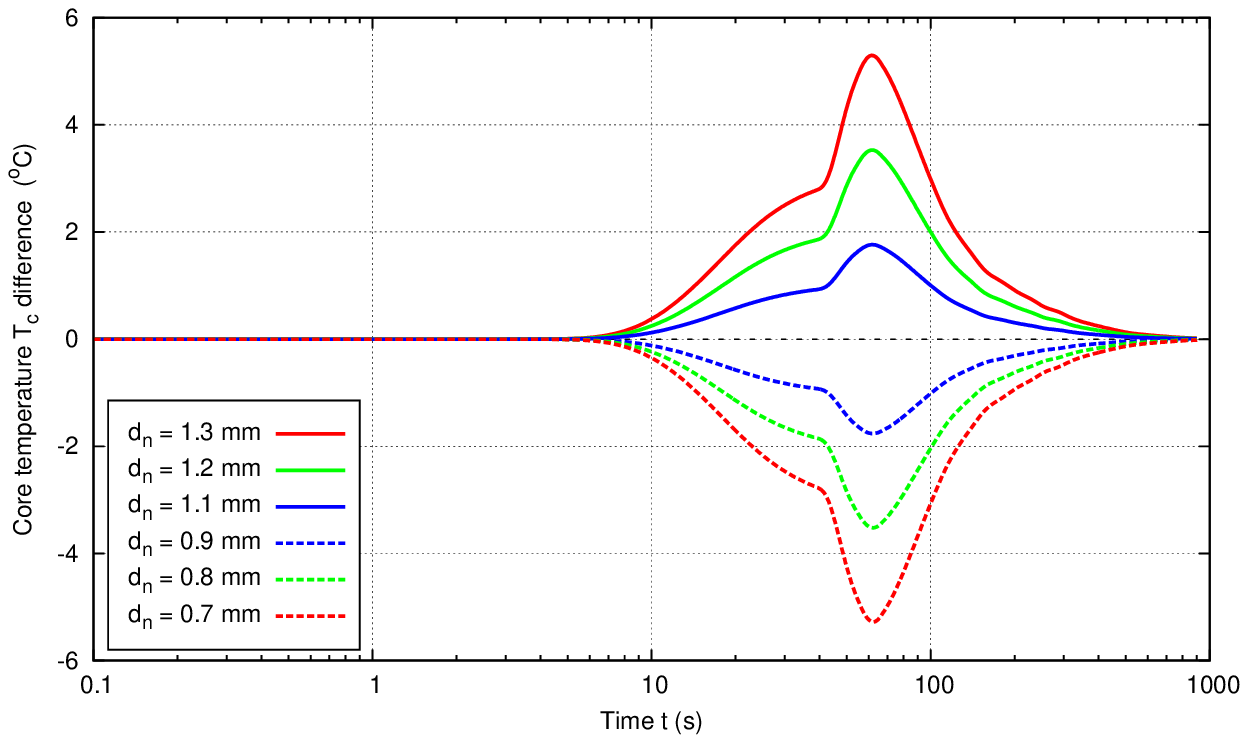}
\end{center}%
\Fig{Oil_Tnp_Tc}%

\vspace*{-3pt}
\noindent
\hbox{Fig.~\ref{Oil_Tnp_Tc} \ }%
	Oil (thermocouple depth variation):
	core temperature $T_c$ differences.
\vspace*{12pt}
%

\begin{center}
	\epsfbox{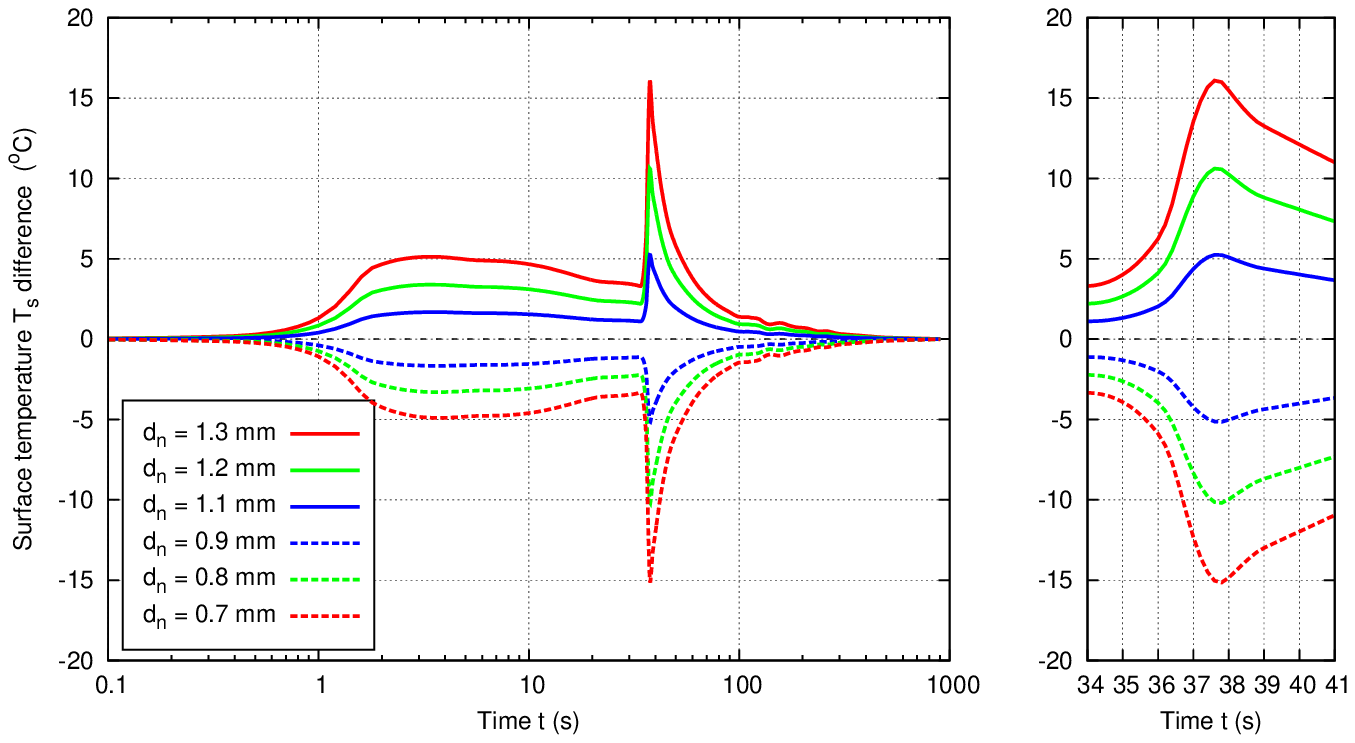}
\end{center}%
\Fig{Oil_Tnp_Ts}%

\vspace*{-3pt}
\noindent
\hbox{Fig.~\ref{Oil_Tnp_Ts} \ }%
	Oil (thermocouple depth variation):
	surface temperature $T_s$ differences.
\vspace*{12pt}
%

\begin{center}
	\epsfbox{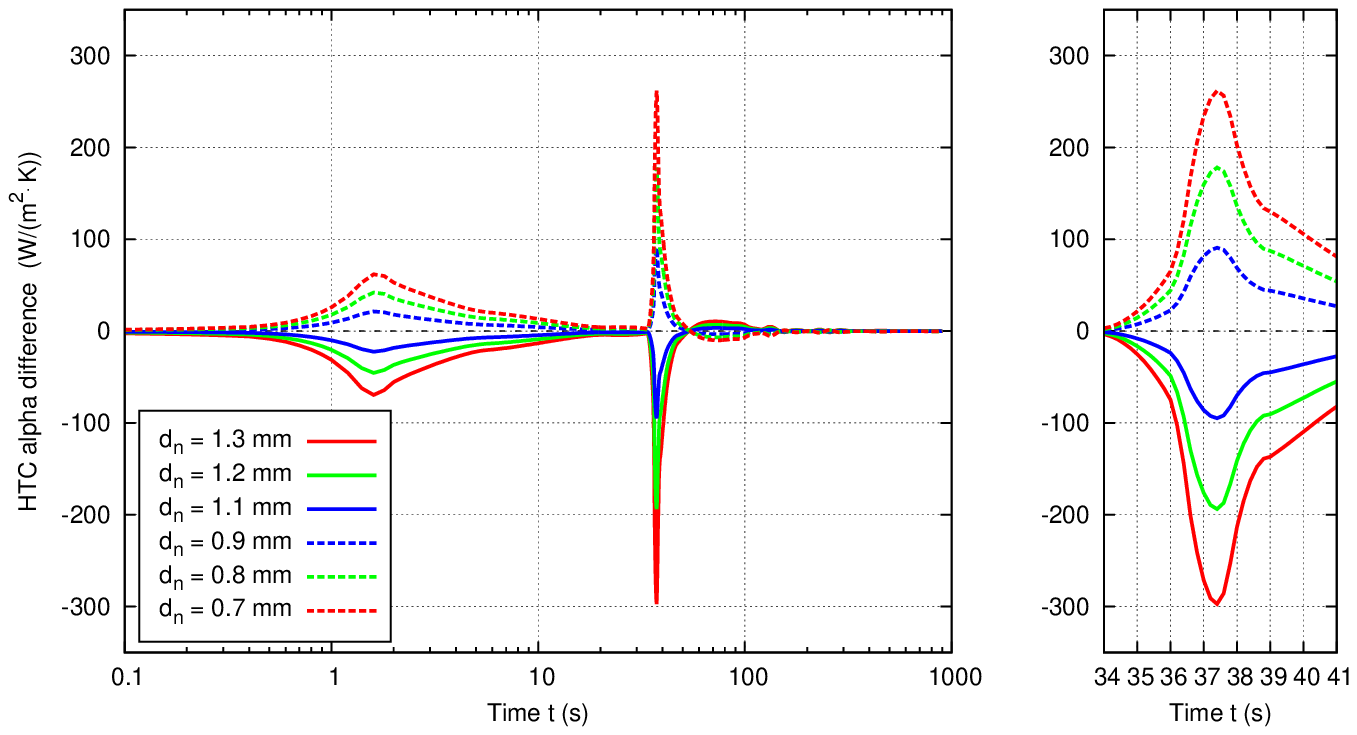}
\end{center}%
\Fig{Oil_Tnp_Alpha}%

\vspace*{-3pt}
\noindent
\hbox{Fig.~\ref{Oil_Tnp_Alpha} \ }%
	Oil (thermocouple depth variation):
	HTC $\alpha$ differences.
\vspace*{12pt}
%

\begin{center}
	\epsfbox{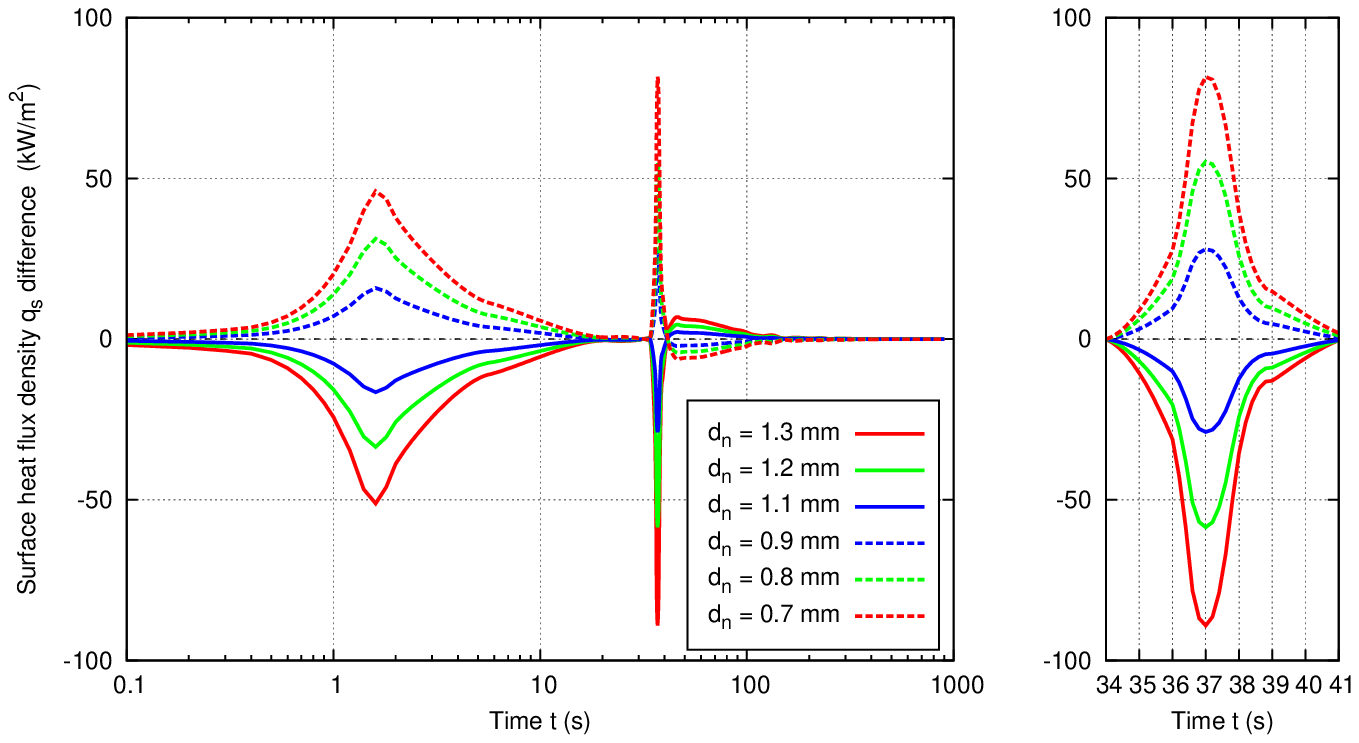}
\end{center}%
\Fig{Oil_Tnp_Flux}%

\vspace*{-3pt}
\noindent
\hbox{Fig.~\ref{Oil_Tnp_Flux} \ }%
	Oil (thermocouple depth variation):
	surface heat flux density $q_s$ differences.
\vspace*{12pt}
%

\begin{center}
	\epsfbox{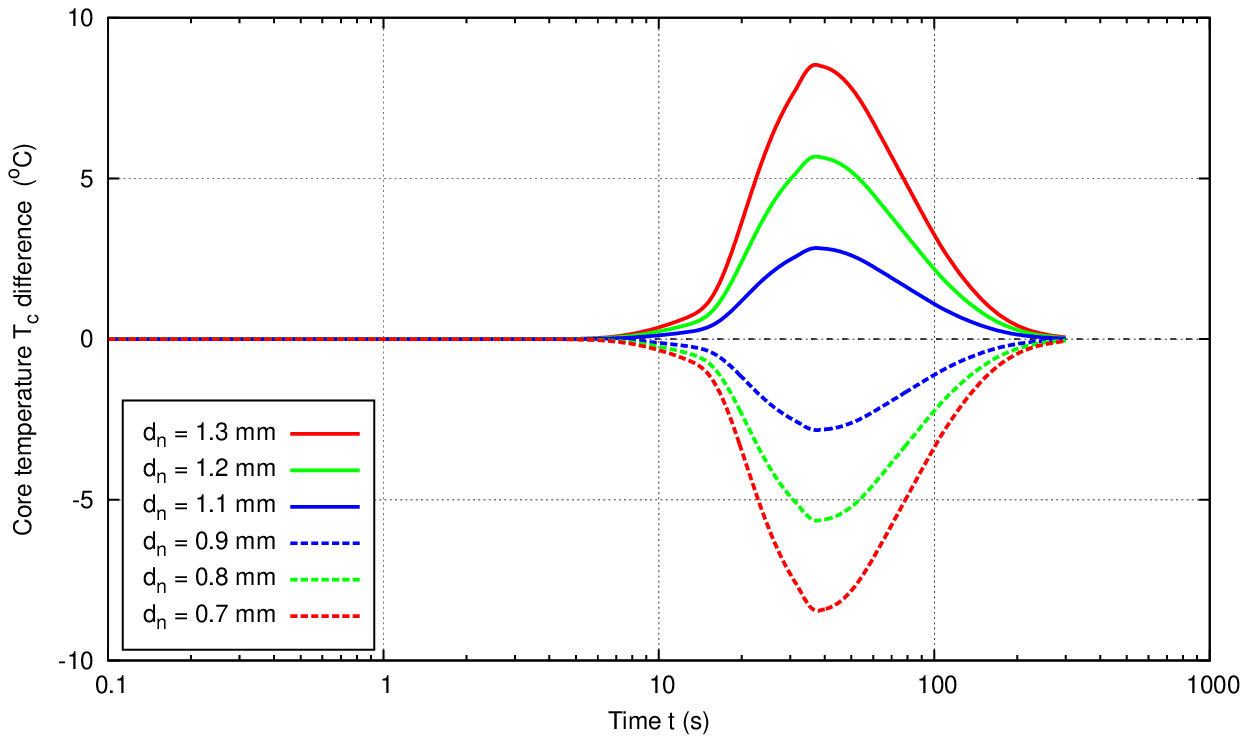}
\end{center}%
\Fig{Wat_Tnp_Tc}%

\vspace*{-3pt}
\noindent
\hbox{Fig.~\ref{Wat_Tnp_Tc} \ }%
	Water (thermocouple depth variation):
	core temperature $T_c$ differences.
\vspace*{12pt}
%

\begin{center}
	\epsfbox{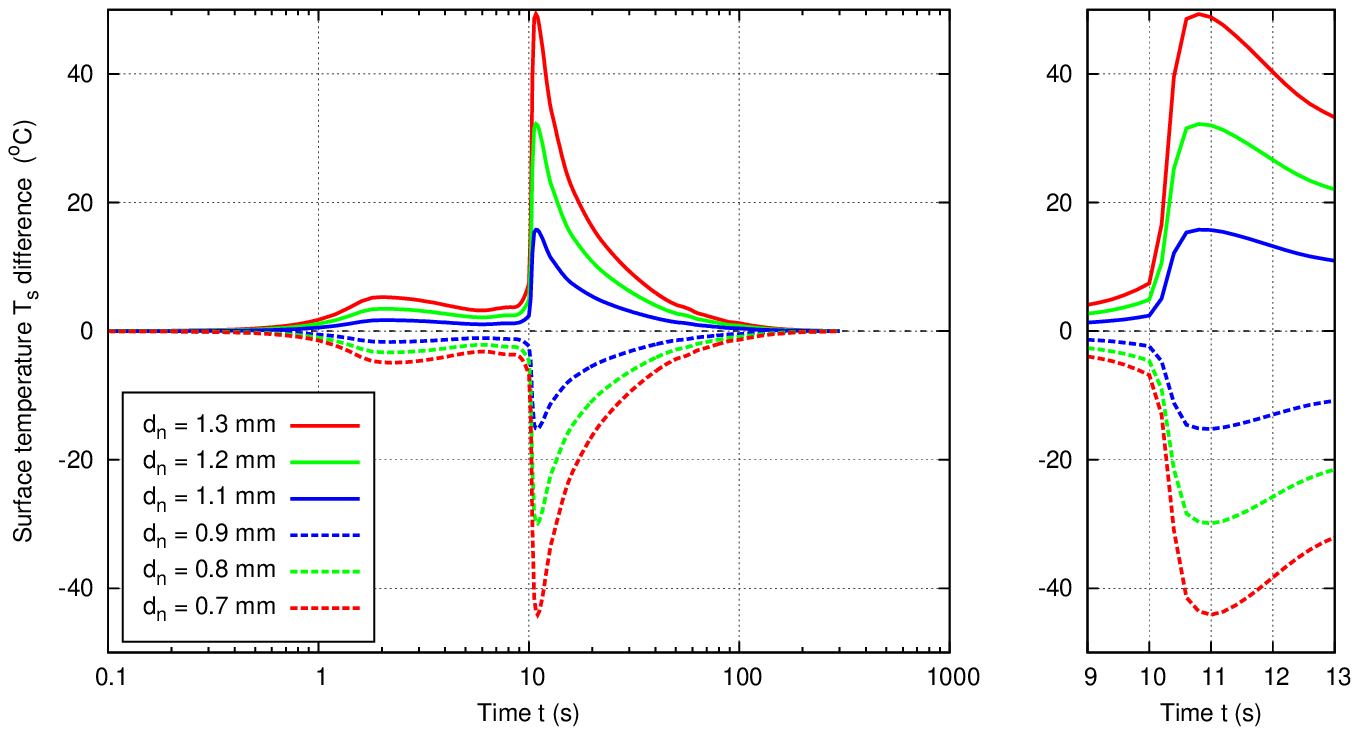}
\end{center}%
\Fig{Wat_Tnp_Ts}%

\vspace*{-3pt}
\noindent
\hbox{Fig.~\ref{Wat_Tnp_Ts} \ }%
	Water (thermocouple depth variation):
	surface temperature $T_s$ differences.
\vspace*{12pt}
%

\begin{center}
	\epsfbox{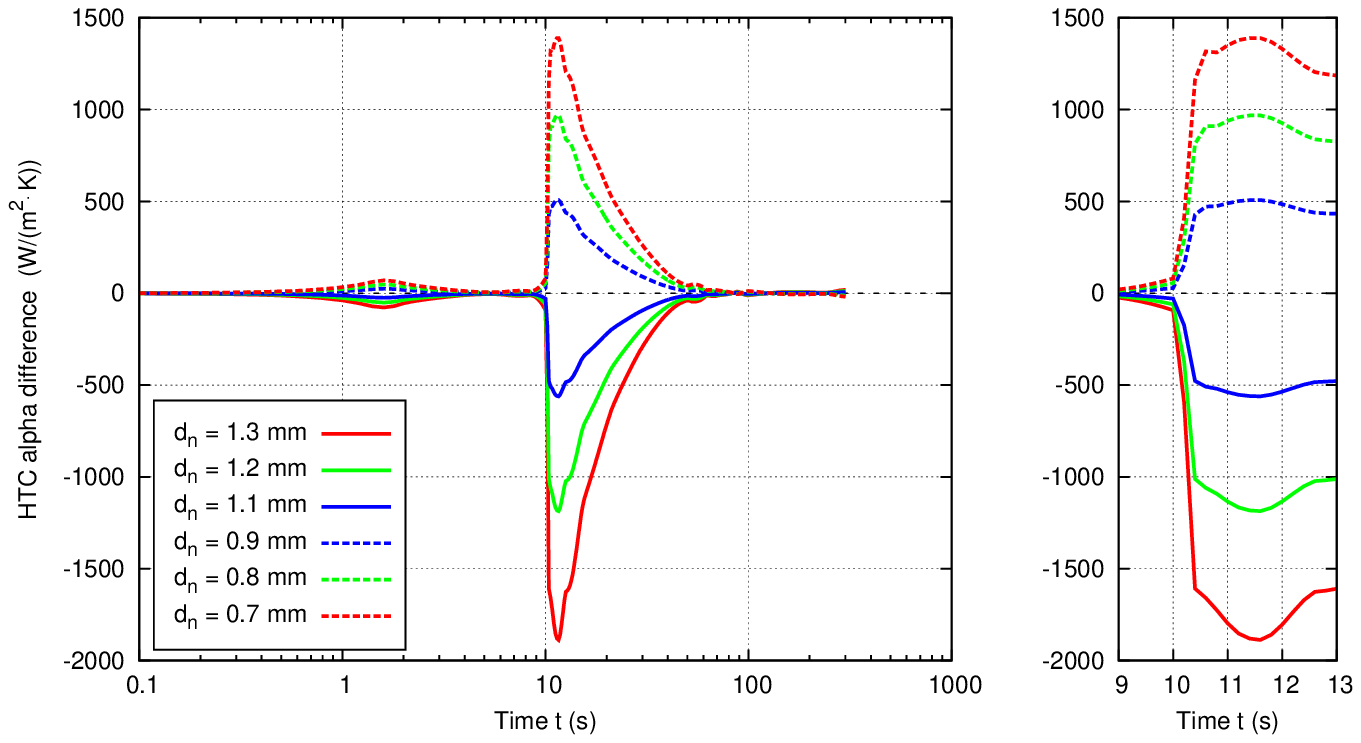}
\end{center}%
\Fig{Wat_Tnp_Alpha}%

\vspace*{-3pt}
\noindent
\hbox{Fig.~\ref{Wat_Tnp_Alpha} \ }%
	Water (thermocouple depth variation):
	HTC $\alpha$ differences.
\vspace*{12pt}
%

\begin{center}
	\epsfbox{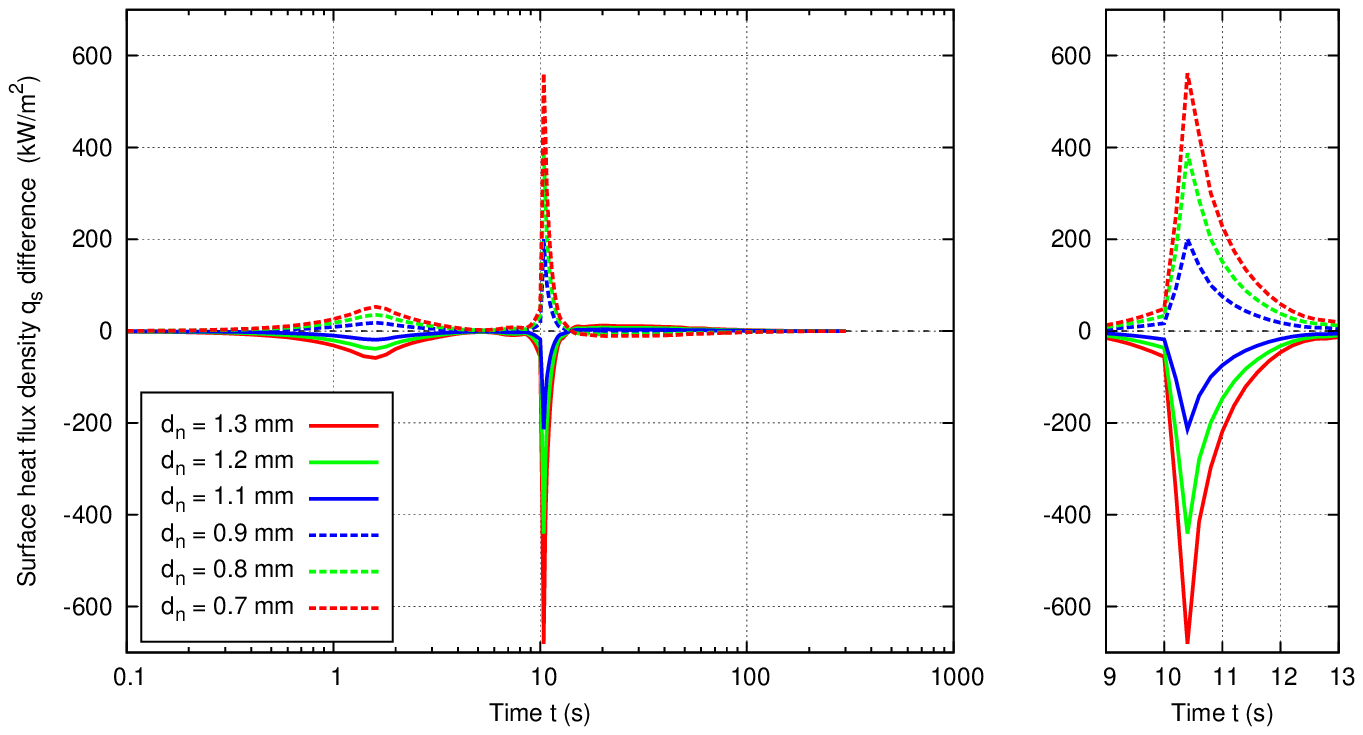}
\end{center}%
\Fig{Wat_Tnp_Flux}%

\vspace*{-3pt}
\noindent
\hbox{Fig.~\ref{Wat_Tnp_Flux} \ }%
	Water (thermocouple depth variation):
	surface heat flux density $q_s$ differences.
\vspace*{12pt}
%

	The magnitudes of differences are higher in water: about
$10 \, {\rm {}^\circ C}$ for the core temperature $T_c$, more than
$40 \, {\rm {}^\circ C}$ for the surface temperature $T_s$, and the HTC
differences range from $-2{,}000$ to $1{,}500 \, {\rm W / (m^2 \cdot K) }$.
The shape of the difference curves is similar, but they last longer, because
of higher magnitudes.

	Because the depth is quite small, and the changes of depth are
relatively large, the resulting differences in the surface conditions, i.e.,
in $T_s$ and $\alpha$, are quite high. On the other hand, the depth changes
are small with respect to the whole radius of the probe, and they happen far
away from the core, so the core temperature $T_c$ is only marginally affected.

%
%
\subsection{Probe diameter}
%
%
	This group of tests may seem odd, since the diameter of the probe
is fixed, and all reference curves are calculated with the diameter
$D = 50 \, {\rm mm}$. However, there are at least two reasons why it is worth
doing. Exactly because the diameter is fixed, someone may be tempted to use
the same output curves for an actual piece of a different size, without being
aware of the errors that might cause (it is has been tried in practice).
From a more positive point of view, this sensitivity test will give at least
some indications about the possibility of using the results for other pieces,
without actually doing the tests with different probes.

	The tests are very similar to the previous group, except that a much
larger step is used, to make the results more interesting for practical
purposes. Six different diameters are tested, with the step of
$5 \, {\rm mm}$. Three of them are smaller, and three are larger than
the standard diameter. Like before, only the diameter $D$ is varied, with
the same input curve $T_n(t)$. Here, of course, it is virtually impossible
that the same temperature would be measured at the same depth
($1 \, {\rm mm}$) in similar quenching conditions, so the results should be
interpreted with this in mind.

	The results of all six tests for quenching in oil are shown in
Figs.~\ref{Oil_Rad_Tc}--\ref{Oil_Rad_Flux}, and the results in water are
shown in Figs.~\ref{Wat_Rad_Tc}--\ref{Wat_Rad_Flux}.
The magnitudes of differences in oil are lower: about
$130 \, {\rm {}^\circ C}$ for the core temperature $T_c$, about
$5 \, {\rm {}^\circ C}$ for the surface temperature $T_s$, and the HTC
differences range from $-400$ to $500 \, {\rm W / (m^2 \cdot K) }$. The shape
of all difference curves is very similar, except that the HTC differences
diminish towards the end of the process, i.e., they behave more regularly
than in water.

\begin{center}
	\epsfbox{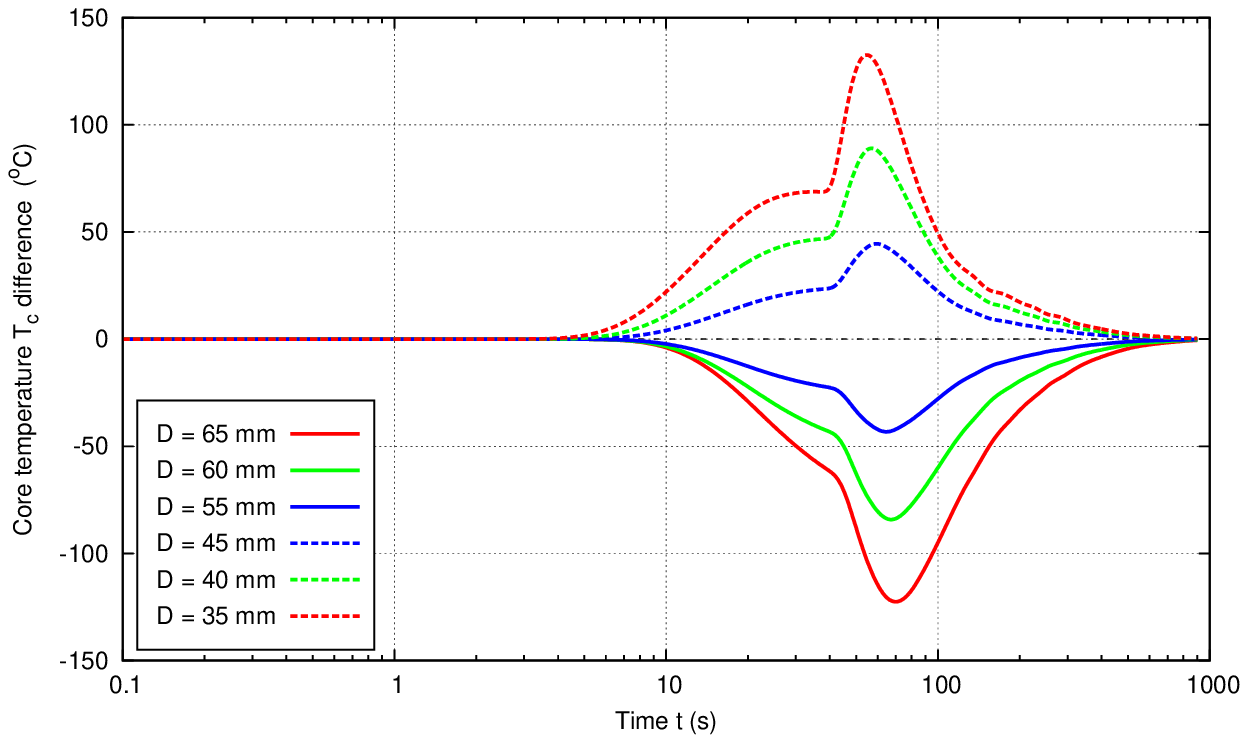}
\end{center}%
\Fig{Oil_Rad_Tc}%

\vspace*{-3pt}
\noindent
\hbox{Fig.~\ref{Oil_Rad_Tc} \ }%
	Oil (diameter variation):
	core temperature $T_c$ differences.
\vspace*{12pt}
%

\begin{center}
	\epsfbox{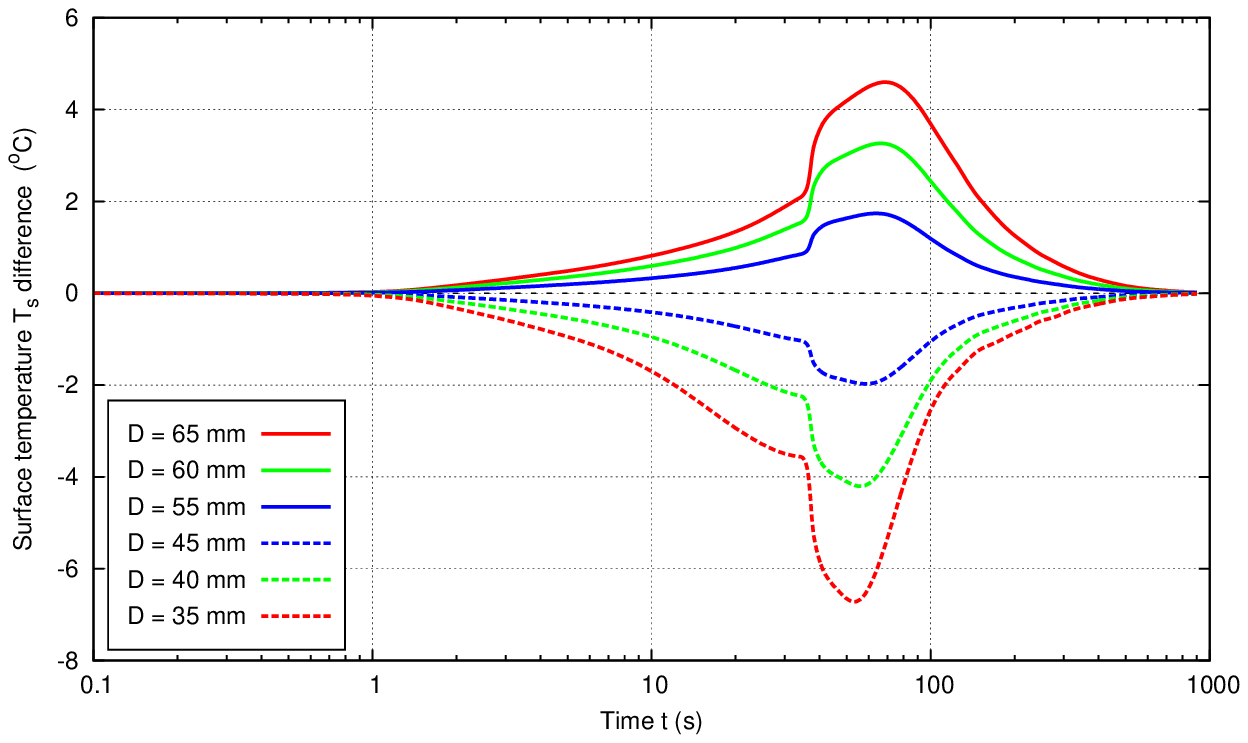}
\end{center}%
\Fig{Oil_Rad_Ts}%

\vspace*{-3pt}
\noindent
\hbox{Fig.~\ref{Oil_Rad_Ts} \ }%
	Oil (diameter variation):
	surface temperature $T_s$ differences.
\vspace*{12pt}
%

\begin{center}
	\epsfbox{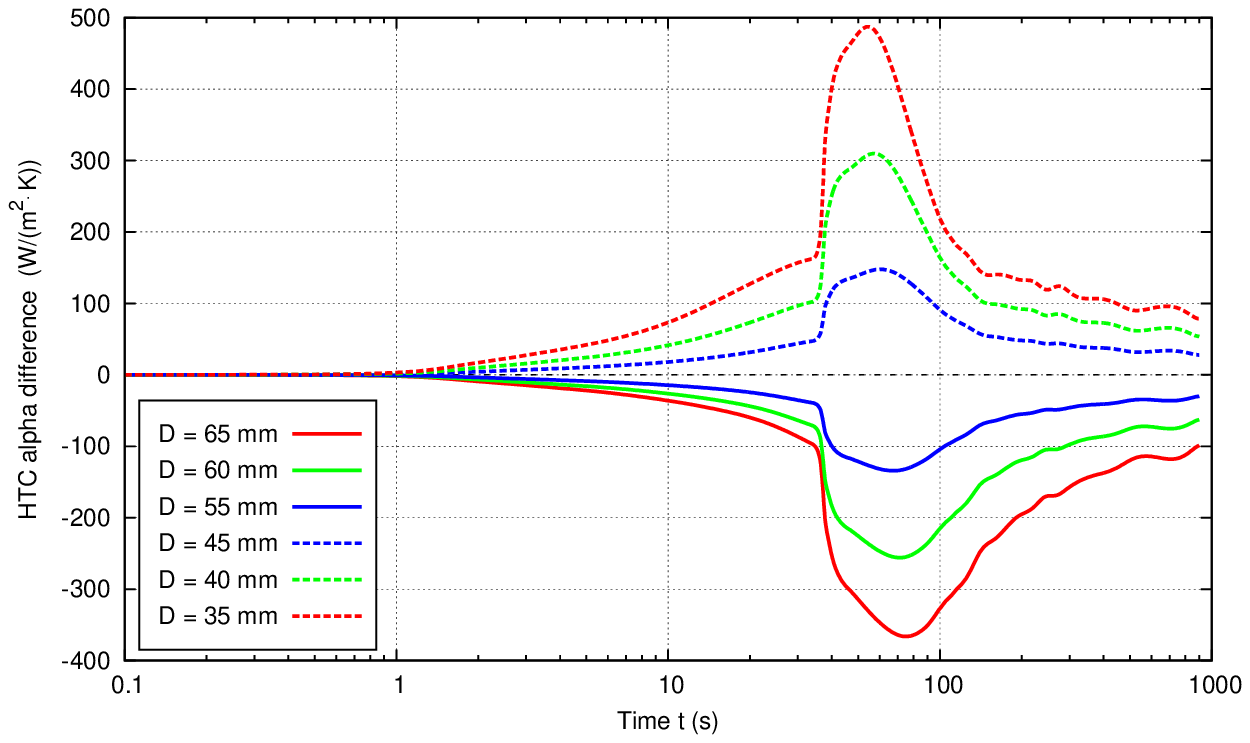}
\end{center}%
\Fig{Oil_Rad_Alpha}%

\vspace*{-3pt}
\noindent
\hbox{Fig.~\ref{Oil_Rad_Alpha} \ }%
	Oil (diameter variation):
	HTC $\alpha$ differences.
\vspace*{12pt}
%

\begin{center}
	\epsfbox{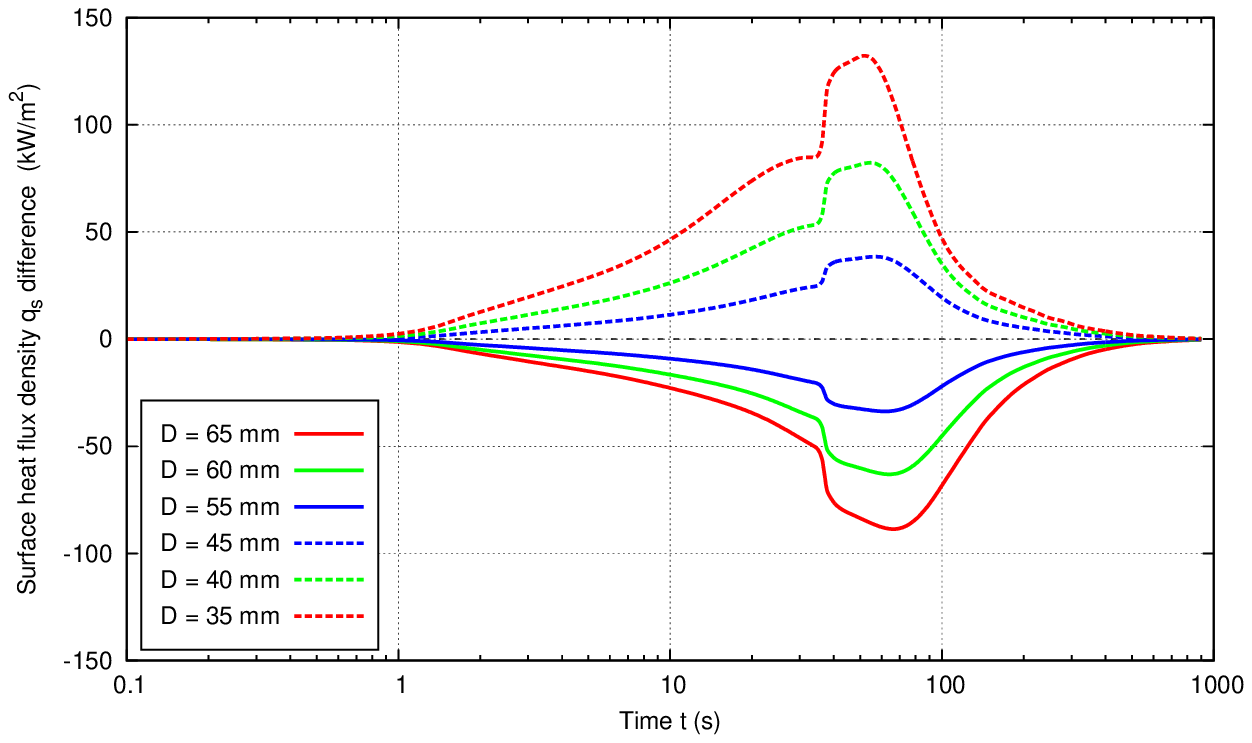}
\end{center}%
\Fig{Oil_Rad_Flux}%

\vspace*{-3pt}
\noindent
\hbox{Fig.~\ref{Oil_Rad_Flux} \ }%
	Oil (diameter variation):
	surface heat flux density $q_s$ differences.
\vspace*{12pt}
%

\begin{center}
	\epsfbox{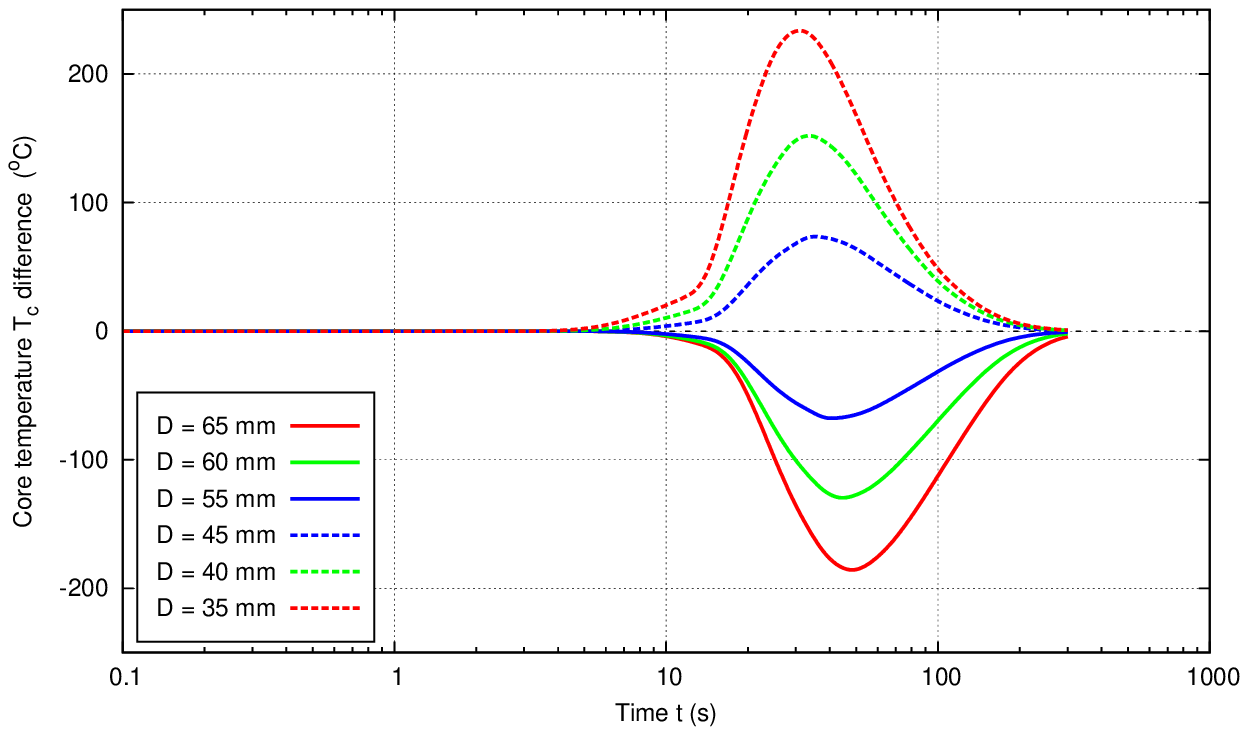}
\end{center}%
\Fig{Wat_Rad_Tc}%

\vspace*{-3pt}
\noindent
\hbox{Fig.~\ref{Wat_Rad_Tc} \ }%
	Water (diameter variation):
	core temperature $T_c$ differences.
\vspace*{12pt}
%

\begin{center}
	\epsfbox{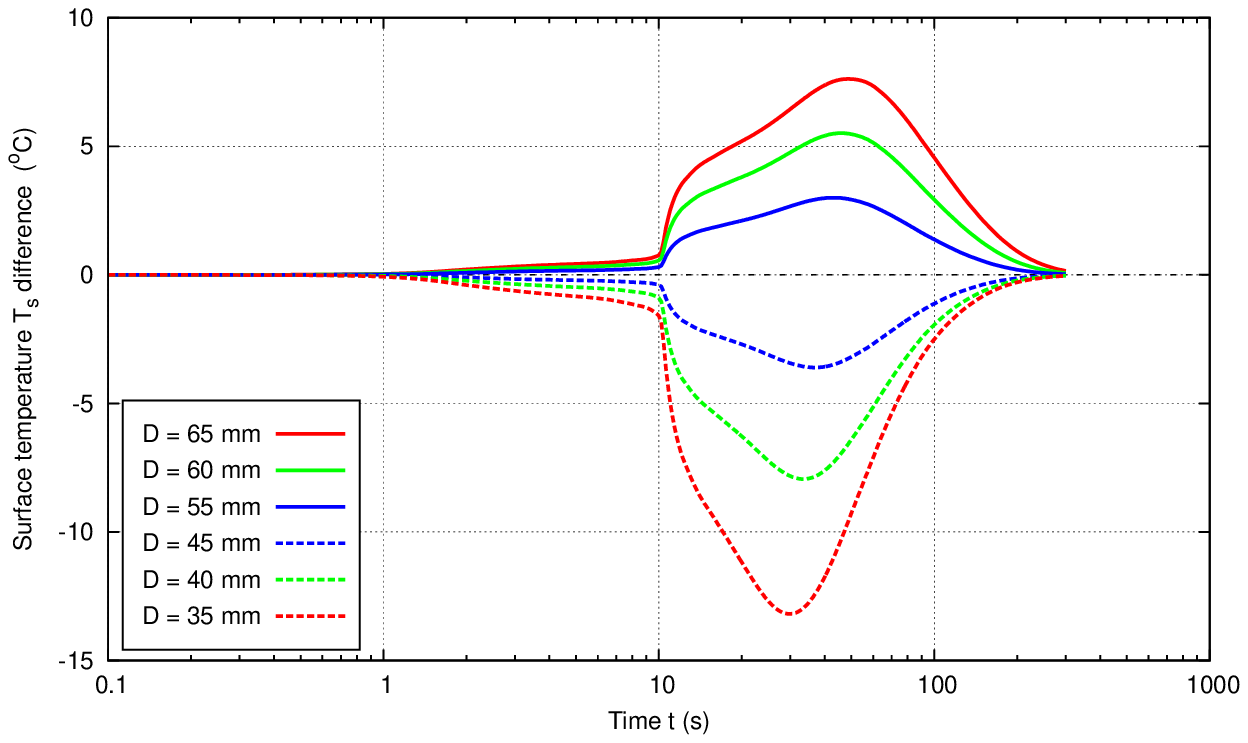}
\end{center}%
\Fig{Wat_Rad_Ts}%

\vspace*{-3pt}
\noindent
\hbox{Fig.~\ref{Wat_Rad_Ts} \ }%
	Water (diameter variation):
	surface temperature $T_s$ differences.
\vspace*{12pt}
%

\begin{center}
	\epsfbox{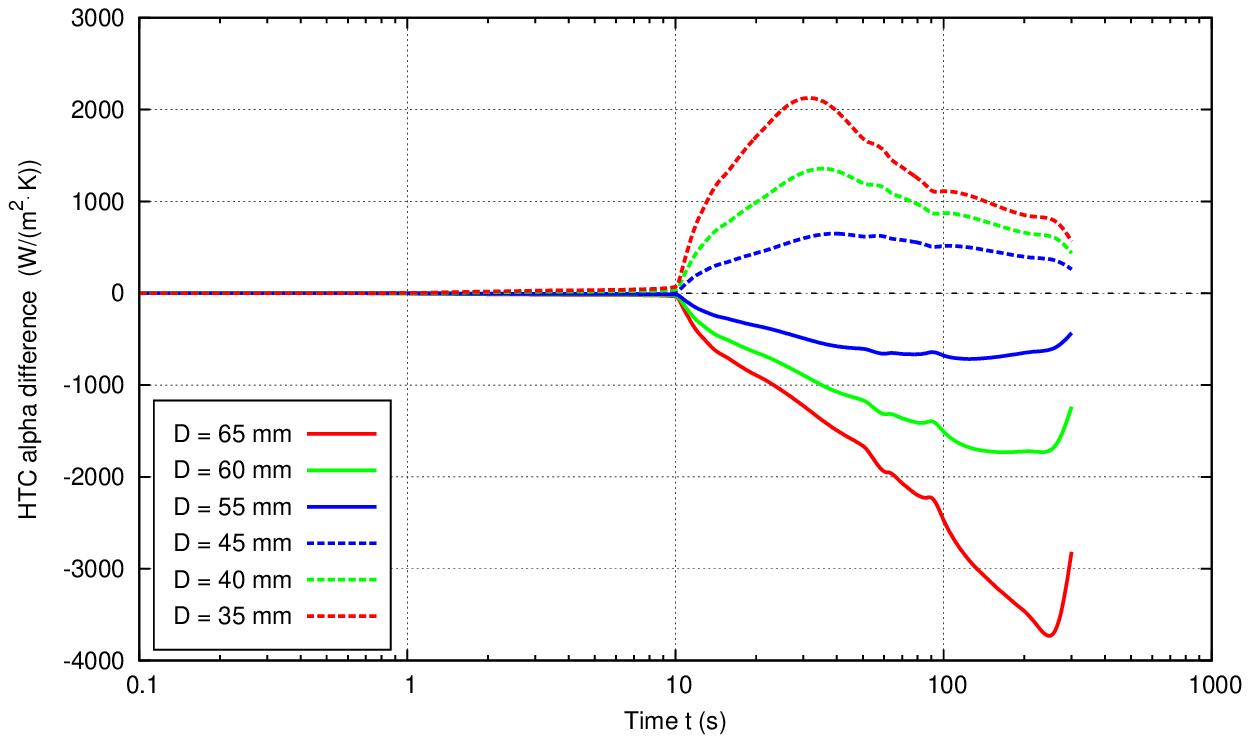}
\end{center}%
\Fig{Wat_Rad_Alpha}%

\vspace*{-3pt}
\noindent
\hbox{Fig.~\ref{Wat_Rad_Alpha} \ }%
	Water (diameter variation):
	HTC $\alpha$ differences.
\vspace*{12pt}
%

\begin{center}
	\epsfbox{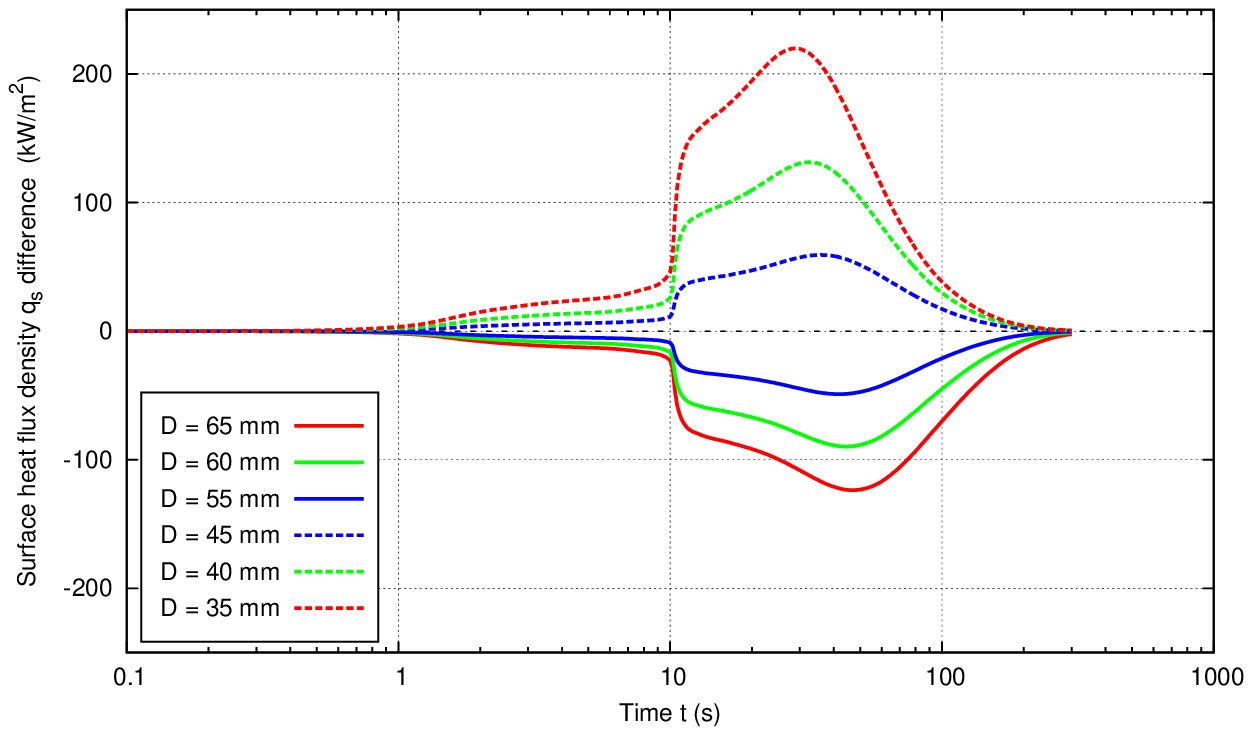}
\end{center}%
\Fig{Wat_Rad_Flux}%

\vspace*{-3pt}
\noindent
\hbox{Fig.~\ref{Wat_Rad_Flux} \ }%
	Water (diameter variation):
	surface heat flux density $q_s$ differences.
\vspace*{12pt}
%

	The results here are quite the opposite from those in the previous
group. Because the depth $d_n$ is fixed, any change of the diameter
effectively changes the distance of the near-surface thermocouple from
the core. Therefore, it is reflected mostly in the core temperature.
The surface conditions are less affected, especially the surface temperature.
The HTC differences are large because of relatively large changes of
the diameter.

	Moreover, the HTC results in Fig.~\ref{Wat_Rad_Alpha} demonstrate
a phenomenon that can occur with the HTC curves calculated in
water---the values at the very end of the quenching process can blow up, with
no apparent limit. Surface heat flux densities in Fig.~\ref{Wat_Rad_Flux}
do not exhibit this behavior, and their values (as well as differences with
respect to the original unperturbed values) nicely tend to zero for high
times. The explanation for high, possibly unbounded, HTC values for high
times comes from Eq~4, where the HTC is calculated as
\begin{equation}
	\alpha = \frac{q_s}{T_s - T_x}.
\label{Eq.6}
\end{equation}
The denominator, i.e., the difference between the surface and the quenchant
temperatures, also tends to zero for high times, and there is no apparent
physical reason that the HTC should be bounded at the time limit.
%
%

	Finally, actual experiments with probes having different diameters,
from $20 \, {\rm mm}$ to $80 \, {\rm mm}$, but of the same design, and in
the same quenching conditions, reported in
\cite{Liscic-Singer-2014,Liscic-Singer-Beitz-2011}, show that the HTC curves
are more similar than the sensitivity results suggest.
For very small diameters, this is not true any more, and the HTC sharply
increases.

%
%
\subsection{Space step}
%
%
	The last two groups of tests are intended to verify the reliability
of the HTC calculation with respect to the main numerical parameters of
the method---the space step and the time step, that are used in
the discretization of the heat conduction problem.

	In the FVM, the whole radius of the probe is divided into a certain
number of finite volumes that determine the space grid, and temperatures are
calculated only at the points of this grid. The grid points are uniformly
spaced, with the space step $h$ (${\rm mm}$), except possibly at the center.
Because the method is based on the solution extension over the interval
$[ R - d_n, R ]$, the near-surface thermocouple position $R - d_n$ is always
taken as the grid point, so that the corresponding temperature is calculated
directly in step~4, without any additional interpolation. Therefore,
the space step $h$ is determined as
\begin{equation}
	h = \frac{d_n}{N_{\rm ext}},
\label{Eq.7}
\end{equation}
where $N_{\rm ext}$ is the prescribed number of the so-called extension
intervals---that many intervals of length $h$ are located between
the near-surface position and the surface. As a result, the central volume
may be twice as wide, to get an integral number of intervals (volumes) over
the whole radius.

	Usually, the results are computed with $N_{\rm ext} = 8$ extension
intervals, or with the space step $h = 0.125 \, {\rm mm}$, giving a total of
$200$ volumes over the radius of the probe. In this group of tests,
the following four values of $N_{\rm ext}$ are taken in Eq~7:
\begin{equation}
	N_{\rm ext} = 1, 2, 4, 10.
\label{Eq.8}
\end{equation}
The space step $h$ varies from $0.1 \, {\rm mm}$ to $1.0 \, {\rm mm}$, and
the total number of volumes varies from $25$ to $250$.

%
%
	The results of all four tests for quenching in oil are shown in
Figs.~\ref{Oil_Nex_Tc}--\ref{Oil_Nex_Flux}, and the results in water are
shown in Figs.~\ref{Wat_Nex_Tc}--\ref{Wat_Nex_Flux}.
	Even with the largest space step, the resulting differences in both
temperatures are very small: less than $0.06 \, {\rm {}^\circ C}$ for oil,
and less than $0.8 \, {\rm {}^\circ C}$ for water.

\begin{center}
	\epsfbox{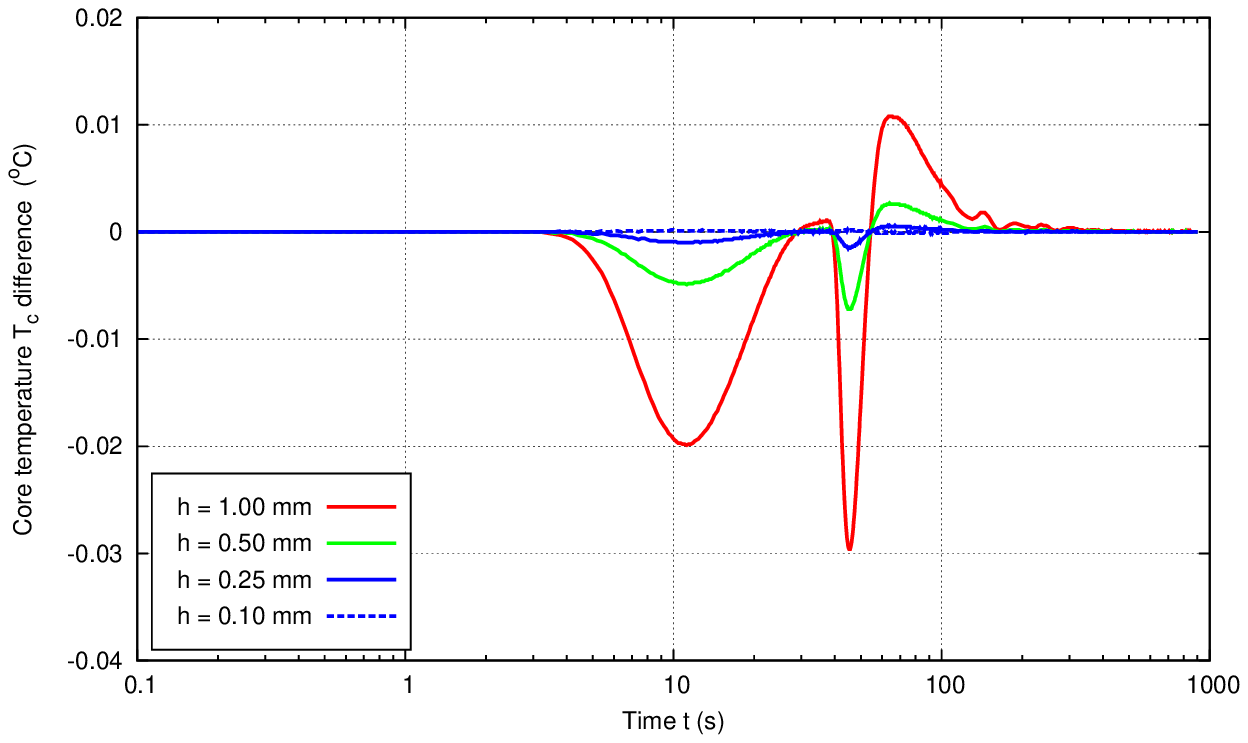}
\end{center}%
\Fig{Oil_Nex_Tc}%

\vspace*{-3pt}
\noindent
\hbox{Fig.~\ref{Oil_Nex_Tc} \ }%
	Oil (space step variation):
	core temperature $T_c$ differences.
\vspace*{12pt}
%

\begin{center}
	\epsfbox{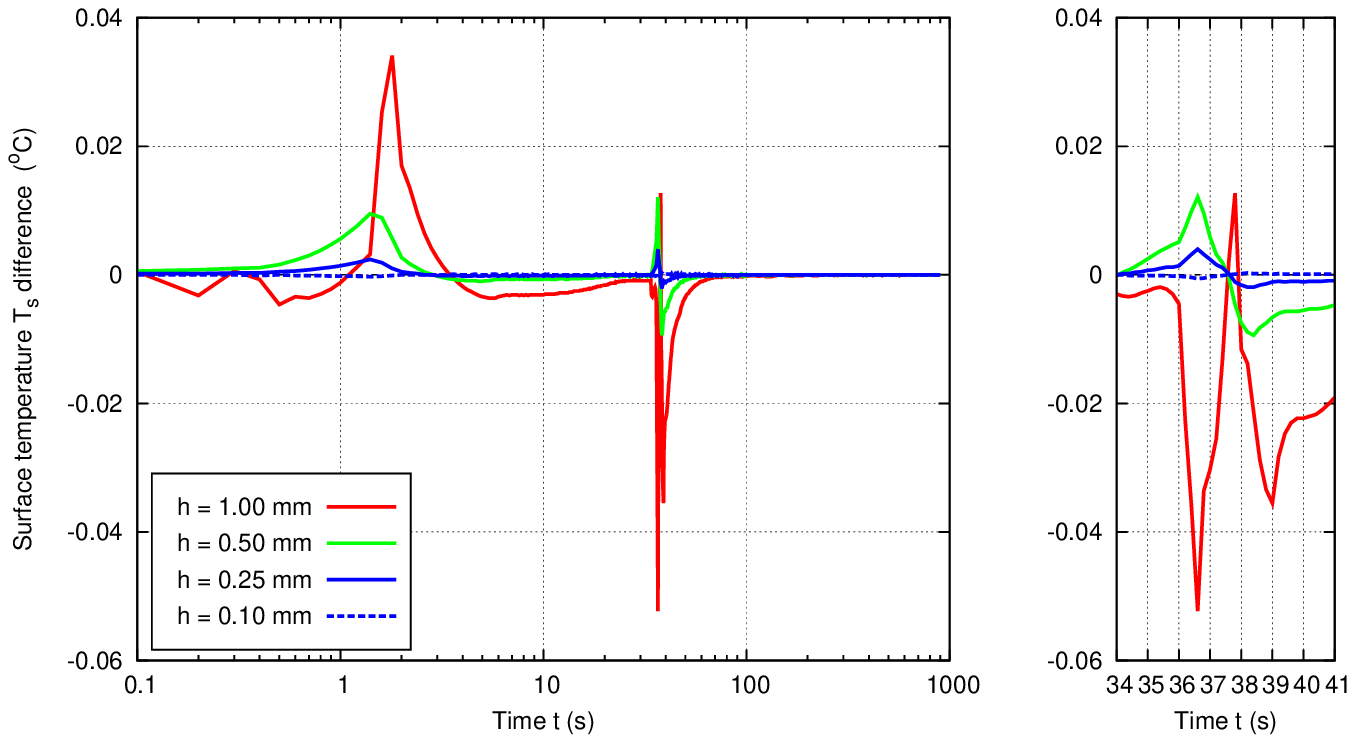}
\end{center}%
\Fig{Oil_Nex_Ts}%

\vspace*{-3pt}
\noindent
\hbox{Fig.~\ref{Oil_Nex_Ts} \ }%
	Oil (space step variation):
	surface temperature $T_s$ differences.
\vspace*{12pt}
%

\begin{center}
	\epsfbox{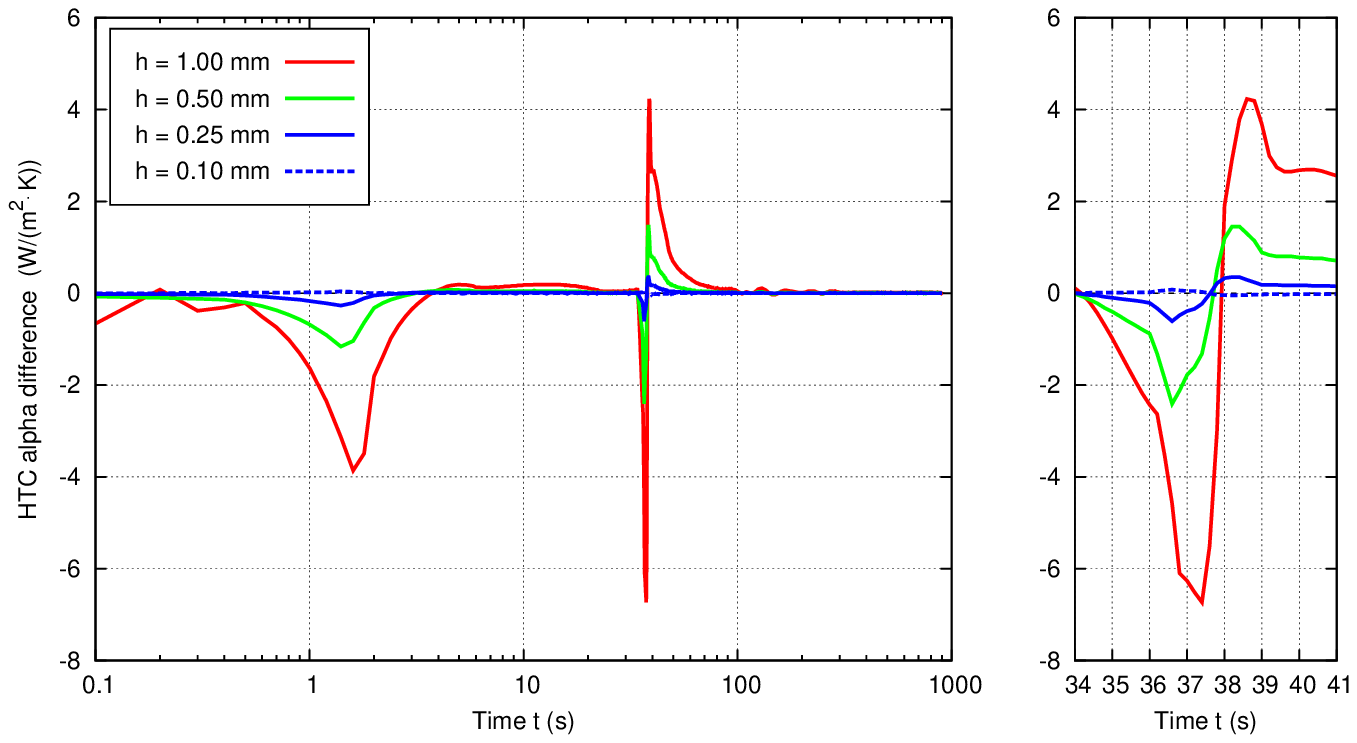}
\end{center}%
\Fig{Oil_Nex_Alpha}%

\vspace*{-3pt}
\noindent
\hbox{Fig.~\ref{Oil_Nex_Alpha} \ }%
	Oil (space step variation):
	HTC $\alpha$ differences.
\vspace*{12pt}
%

\begin{center}
	\epsfbox{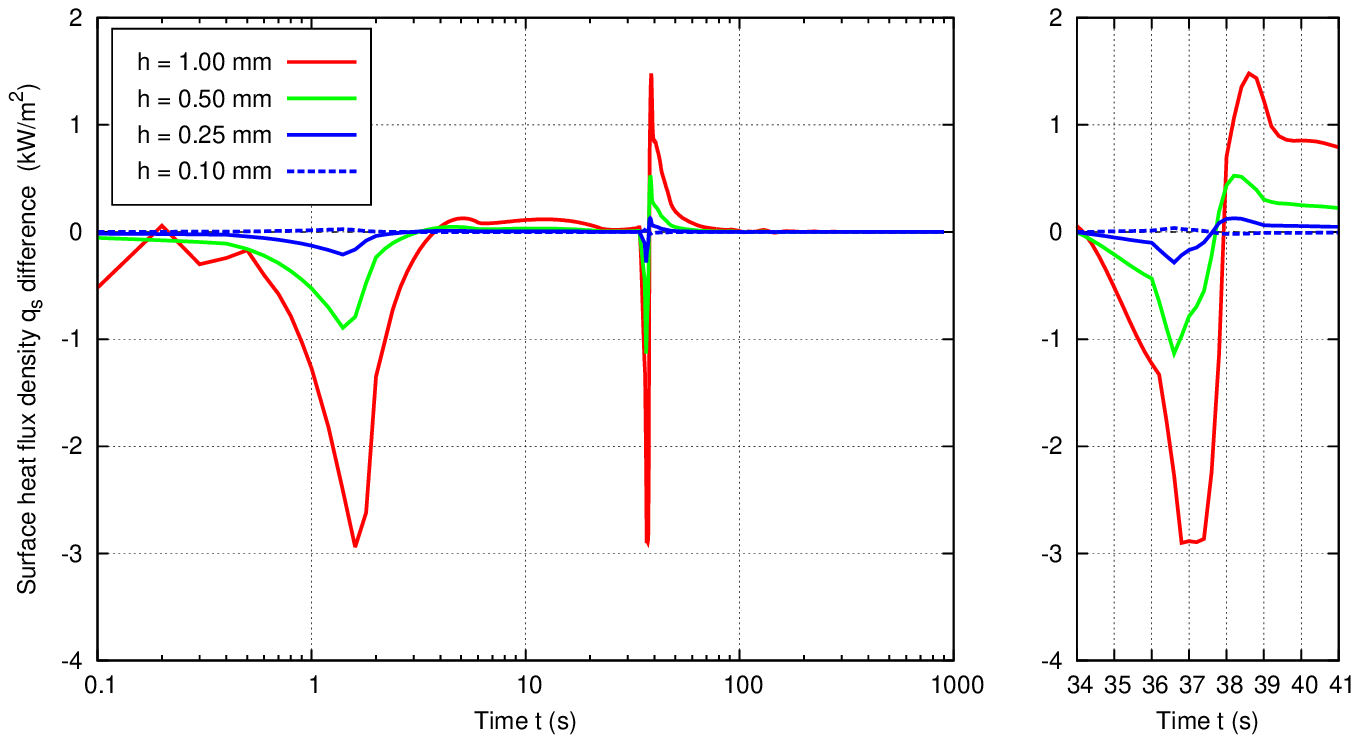}
\end{center}%
\Fig{Oil_Nex_Flux}%

\vspace*{-3pt}
\noindent
\hbox{Fig.~\ref{Oil_Nex_Flux} \ }%
	Oil (space step variation):
	surface heat flux density $q_s$ differences.
\vspace*{12pt}
%

\begin{center}
	\epsfbox{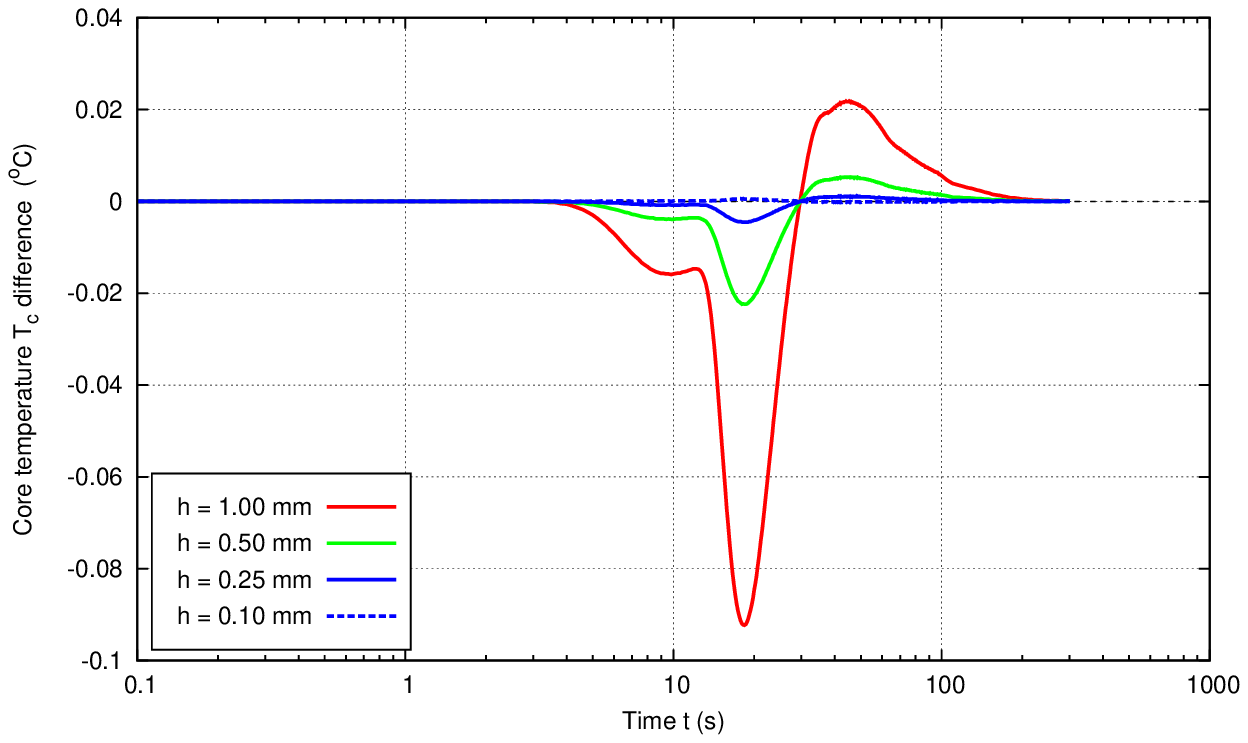}
\end{center}%
\Fig{Wat_Nex_Tc}%

\vspace*{-3pt}
\noindent
\hbox{Fig.~\ref{Wat_Nex_Tc} \ }%
	Water (space step variation):
	core temperature $T_c$ differences.
\vspace*{12pt}
%

\begin{center}
	\epsfbox{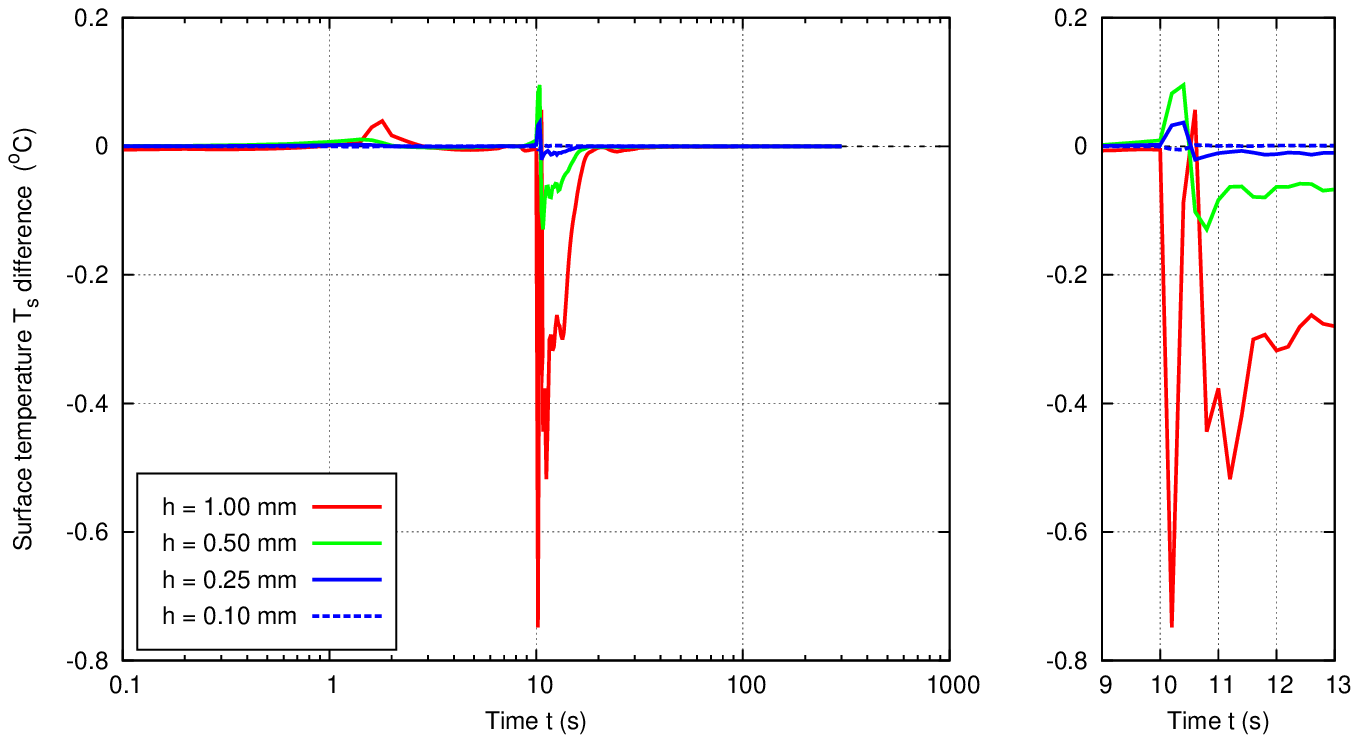}
\end{center}%
\Fig{Wat_Nex_Ts}%

\vspace*{-3pt}
\noindent
\hbox{Fig.~\ref{Wat_Nex_Ts} \ }%
	Water (space step variation):
	surface temperature $T_s$ differences.
\vspace*{12pt}
%

\begin{center}
	\epsfbox{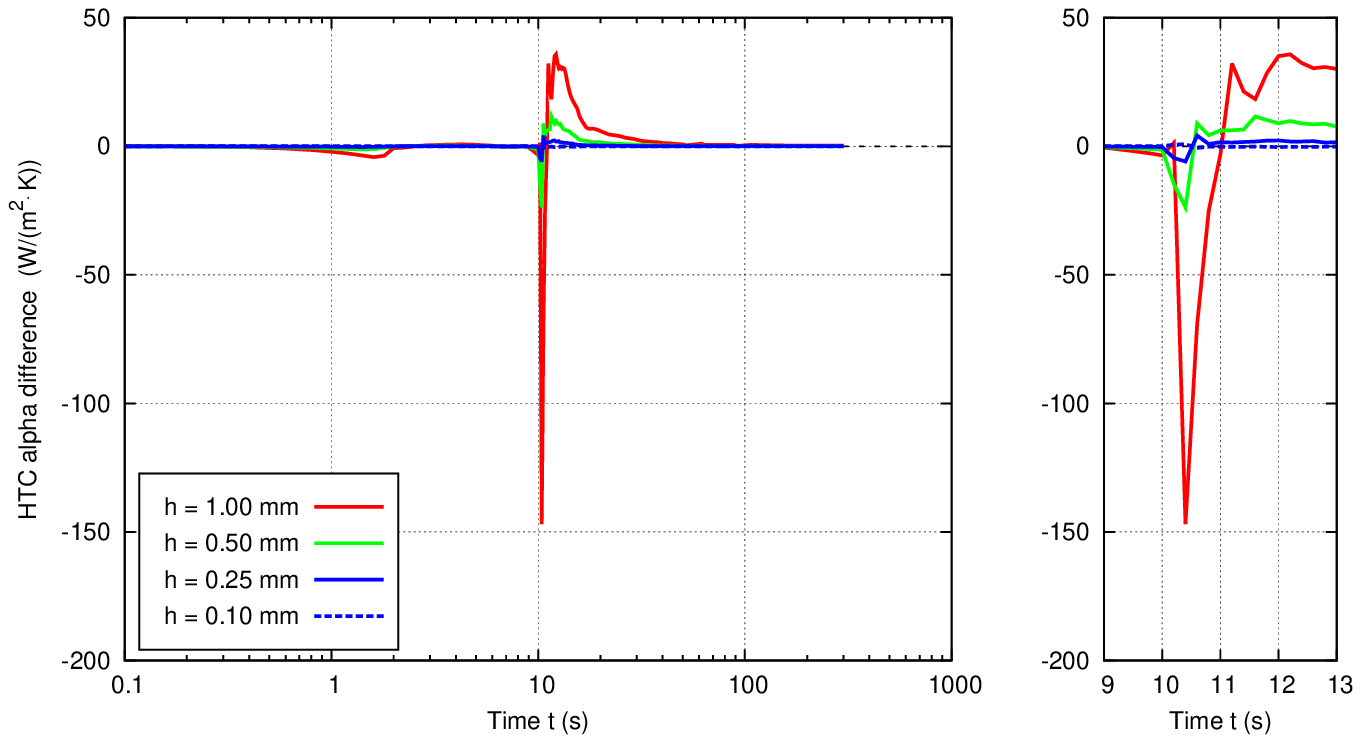}
\end{center}%
\Fig{Wat_Nex_Alpha}%

\vspace*{-3pt}
\noindent
\hbox{Fig.~\ref{Wat_Nex_Alpha} \ }%
	Water (space step variation):
	HTC $\alpha$ differences.
\vspace*{12pt}
%

\begin{center}
	\epsfbox{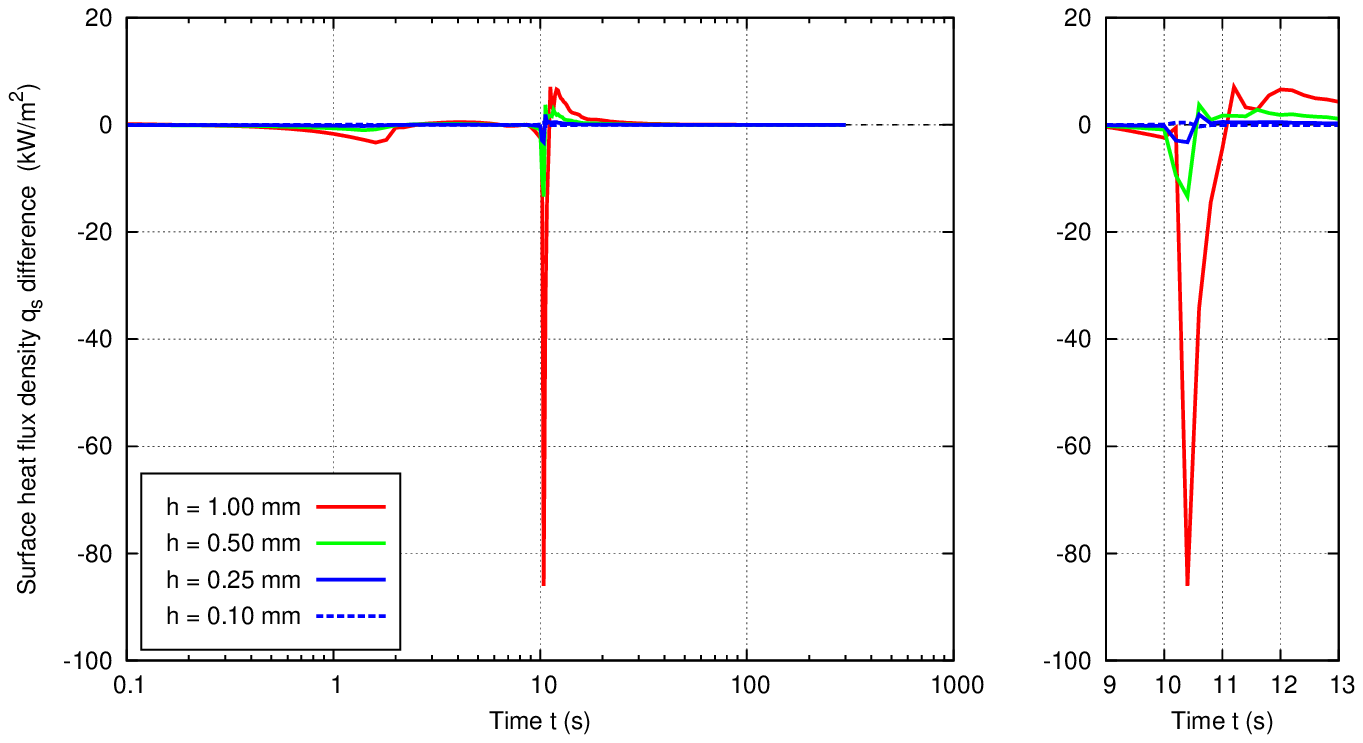}
\end{center}%
\Fig{Wat_Nex_Flux}%

\vspace*{-3pt}
\noindent
\hbox{Fig.~\ref{Wat_Nex_Flux} \ }%
	Water (space step variation):
	surface heat flux density $q_s$ differences.
\vspace*{12pt}
%

	High differences in the HTCs are strictly confined to the intensive
cooling period, and large peaks are obtained only when a single extension
interval is used. In all other cases, the differences may be considered as
negligible, for all practical purposes.

	The method becomes slightly less stable with a very large number
of extension intervals ($N_{\rm ext} \geq 12$), and the usual value of $8$
extension intervals turns out to be a good choice.

%
%
\subsection{Time step}
%
%
	The overall space accuracy of the method is reasonably high,
proportional to $h^2$. In contrast to that, because it is based on
the implicit advance in time, its accuracy with respect to time is quite
low, directly proportional to the time step. So, despite the fact that
large time steps can be used without a sacrifice in stability, they
still have to be chosen with some care to preserve the accuracy of
the computed solution.

	In order to speed up the calculation, different time steps $\tau_i$
are used in different periods of the quenching process. The computation
begins with the finest time step, denoted by $\tau$ (${\rm s}$), that is
used throughout the most intensive part of the process. For the standard
probe, this time step is used for the first $100$ seconds. Later on, when
the temperature drops, much coarser time steps are used.

	The standard value for the finest time step is
$\tau = 0.02 \, {\rm s}$, in accordance with the frequency of measurements
gathered by the probe. All the results presented thus far, have been computed
with this time step. In this group of tests, the following values of
the finest time step are used:
\begin{equation}
	\tau = 0.01, \, 0.025, \, 0.05, \, 0.1 \, {\rm s}.
\label{Eq.9}
\end{equation}
The first one is finer than the standard step, while the last one is five
times coarser.

	The results of these tests for quenching in oil are shown in
Figs.~\ref{Oil_Tau_Tc}--\ref{Oil_Tau_Flux}, and the results in water are
shown in Figs.~\ref{Wat_Tau_Tc}--\ref{Wat_Tau_Flux}.
	Similarly to the previous group, even with the largest time step,
the resulting differences in both temperatures are quite small: about
$0.6 \, {\rm {}^\circ C}$ for oil, and less than $7.0 \, {\rm {}^\circ C}$
for water.

\begin{center}
	\epsfbox{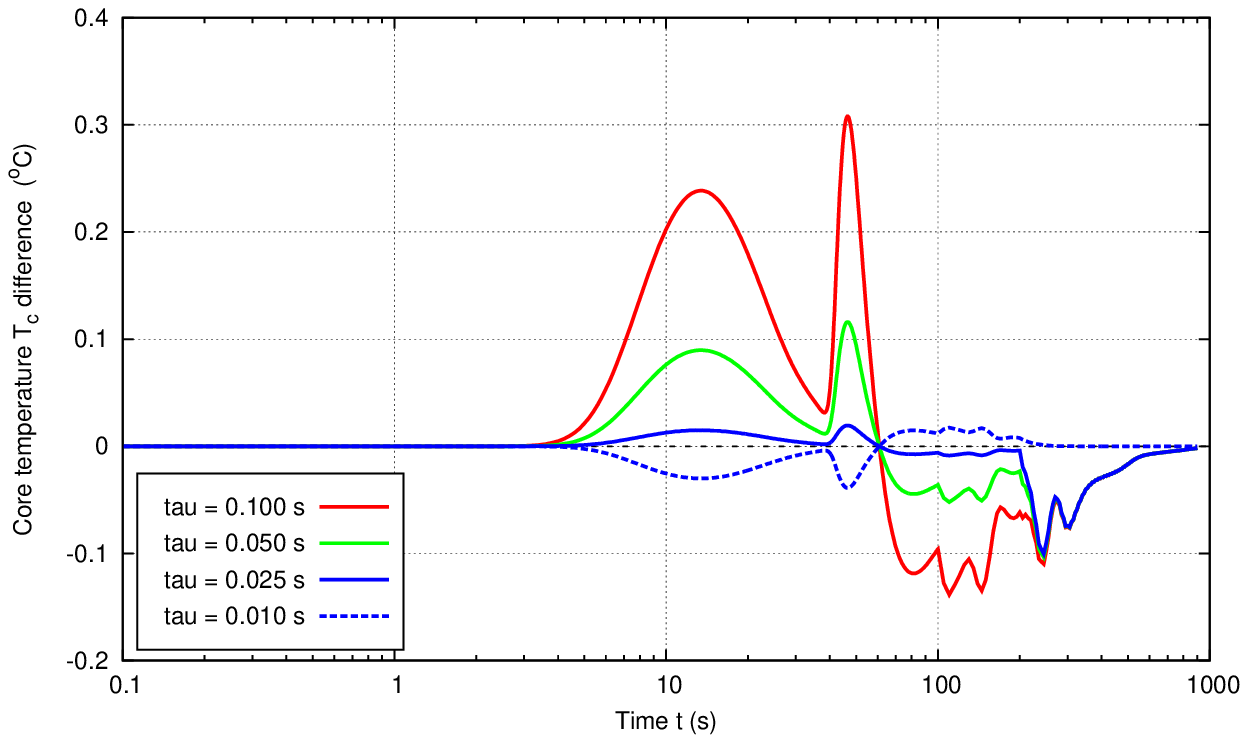}
\end{center}%
\Fig{Oil_Tau_Tc}%

\vspace*{-3pt}
\noindent
\hbox{Fig.~\ref{Oil_Tau_Tc} \ }%
	Oil (time step variation):
	core temperature $T_c$ differences.
\vspace*{12pt}
%

\begin{center}
	\epsfbox{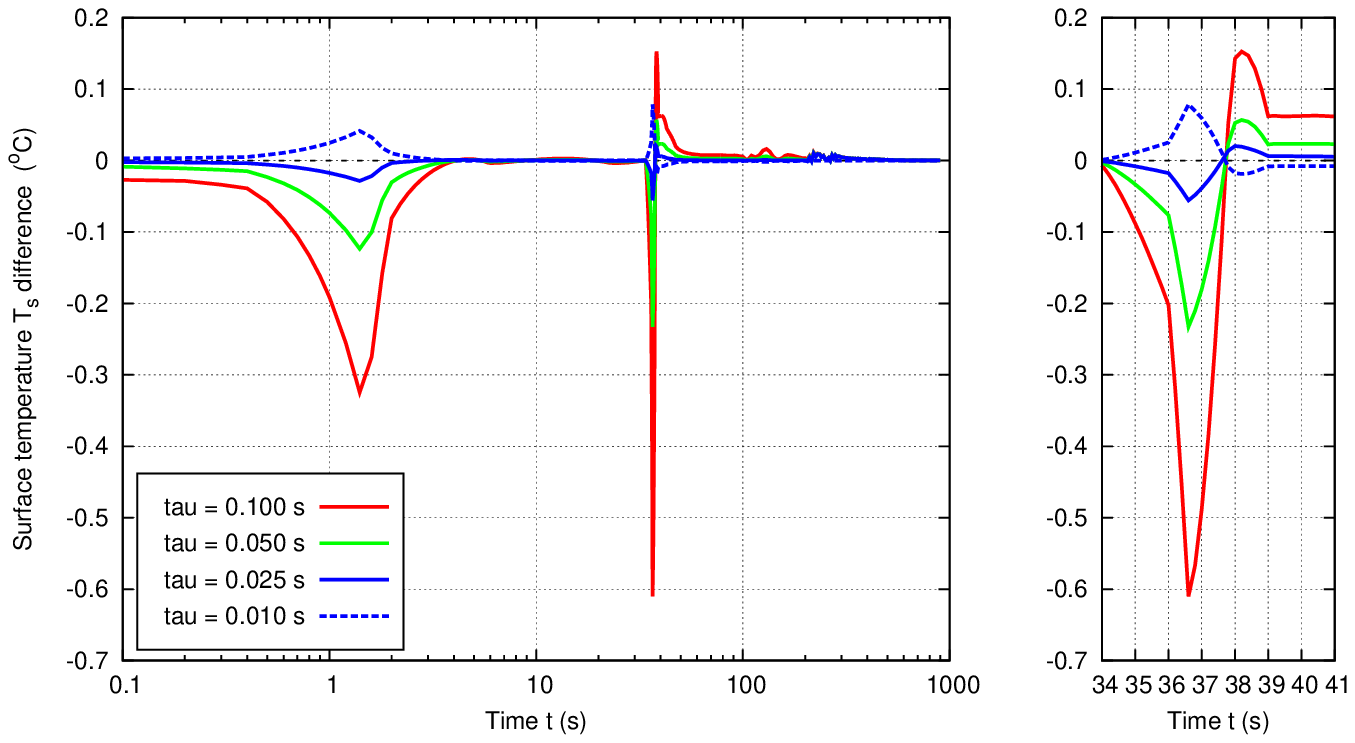}
\end{center}%
\Fig{Oil_Tau_Ts}%

\vspace*{-3pt}
\noindent
\hbox{Fig.~\ref{Oil_Tau_Ts} \ }%
	Oil (time step variation):
	surface temperature $T_s$ differences.
\vspace*{12pt}
%

\begin{center}
	\epsfbox{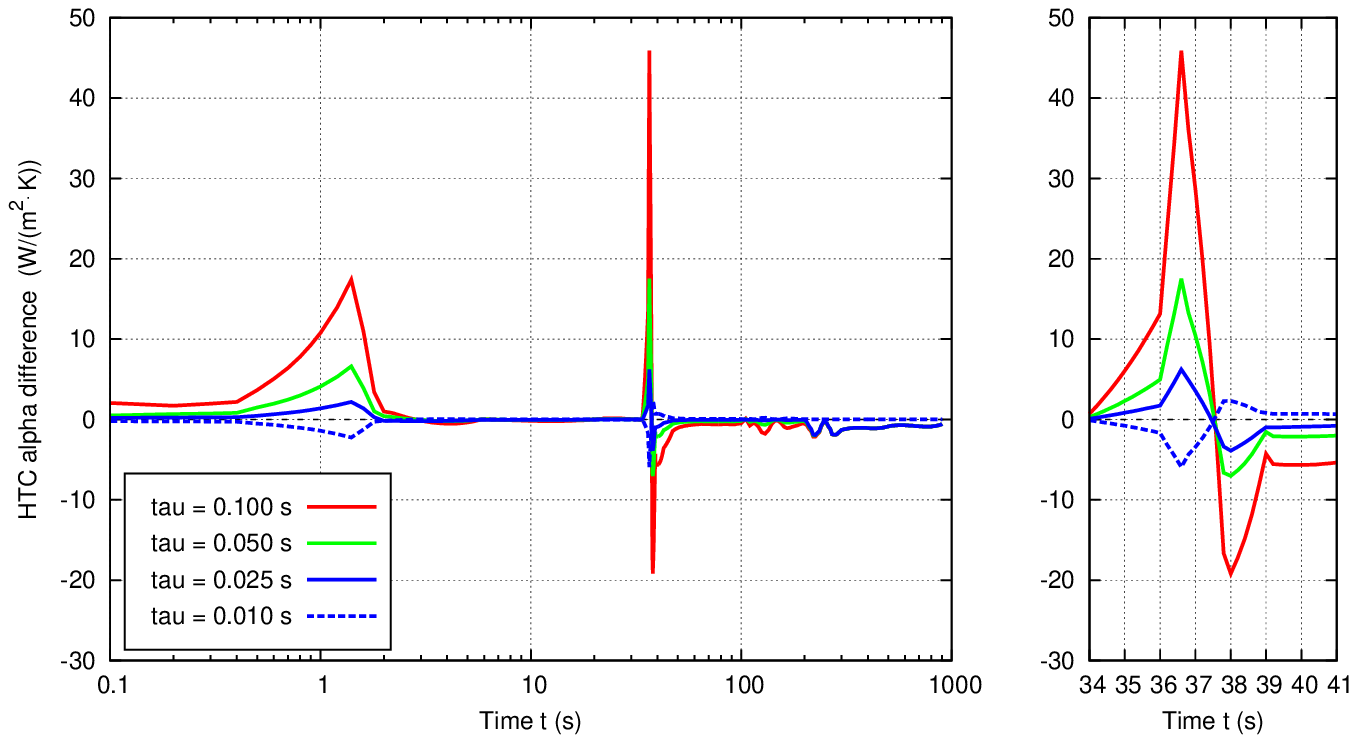}
\end{center}%
\Fig{Oil_Tau_Alpha}%

\vspace*{-3pt}
\noindent
\hbox{Fig.~\ref{Oil_Tau_Alpha} \ }%
	Oil (time step variation):
	HTC $\alpha$ differences.
\vspace*{12pt}
%

\begin{center}
	\epsfbox{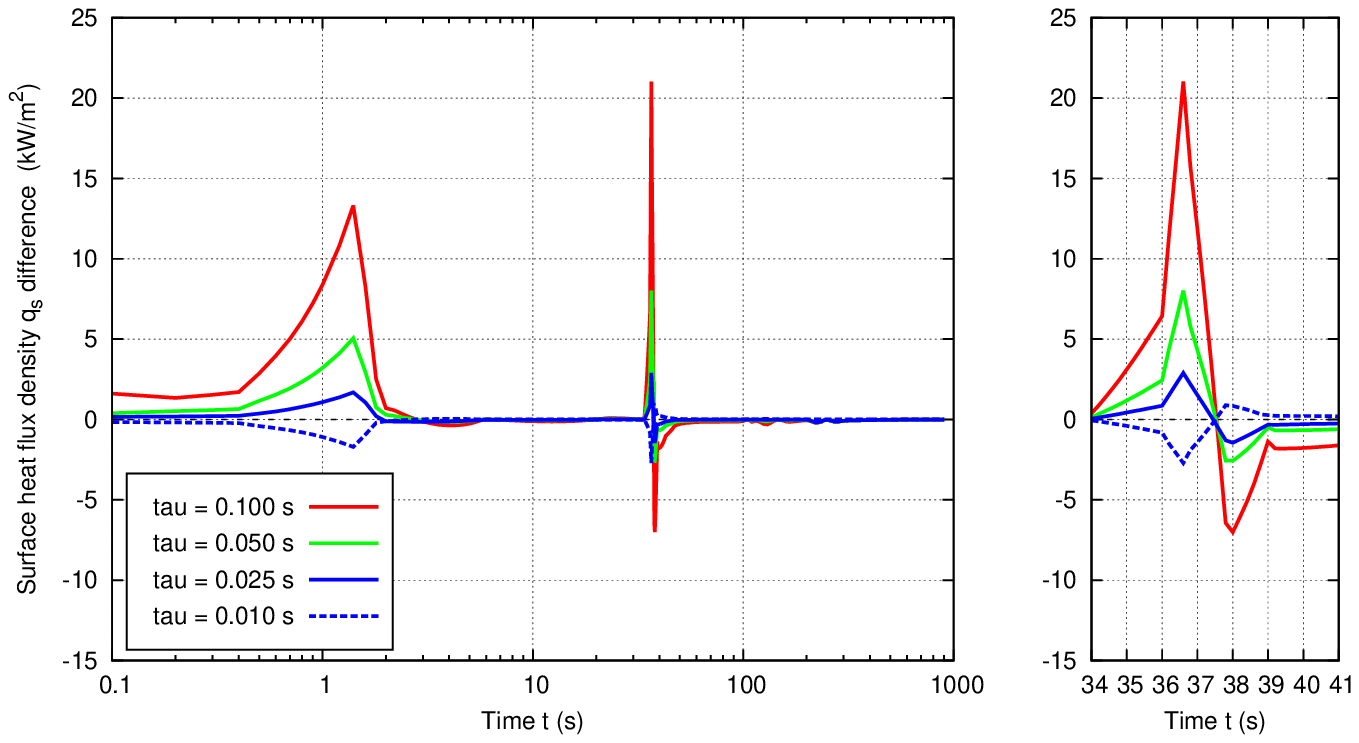}
\end{center}%
\Fig{Oil_Tau_Flux}%

\vspace*{-3pt}
\noindent
\hbox{Fig.~\ref{Oil_Tau_Flux} \ }%
	Oil (time step variation):
	surface heat flux density $q_s$ differences.
\vspace*{12pt}
%

\begin{center}
	\epsfbox{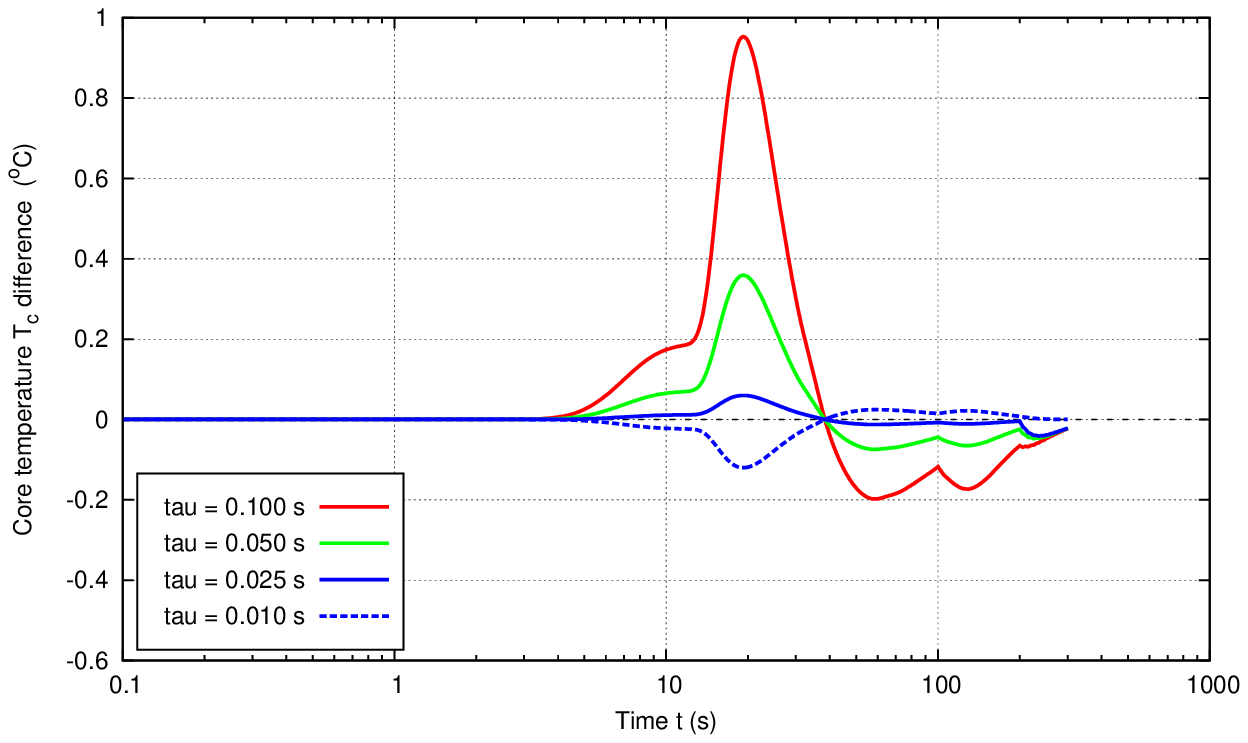}
\end{center}%
\Fig{Wat_Tau_Tc}%

\vspace*{-3pt}
\noindent
\hbox{Fig.~\ref{Wat_Tau_Tc} \ }%
	Water (time step variation):
	core temperature $T_c$ differences.
\vspace*{12pt}
%

\begin{center}
	\epsfbox{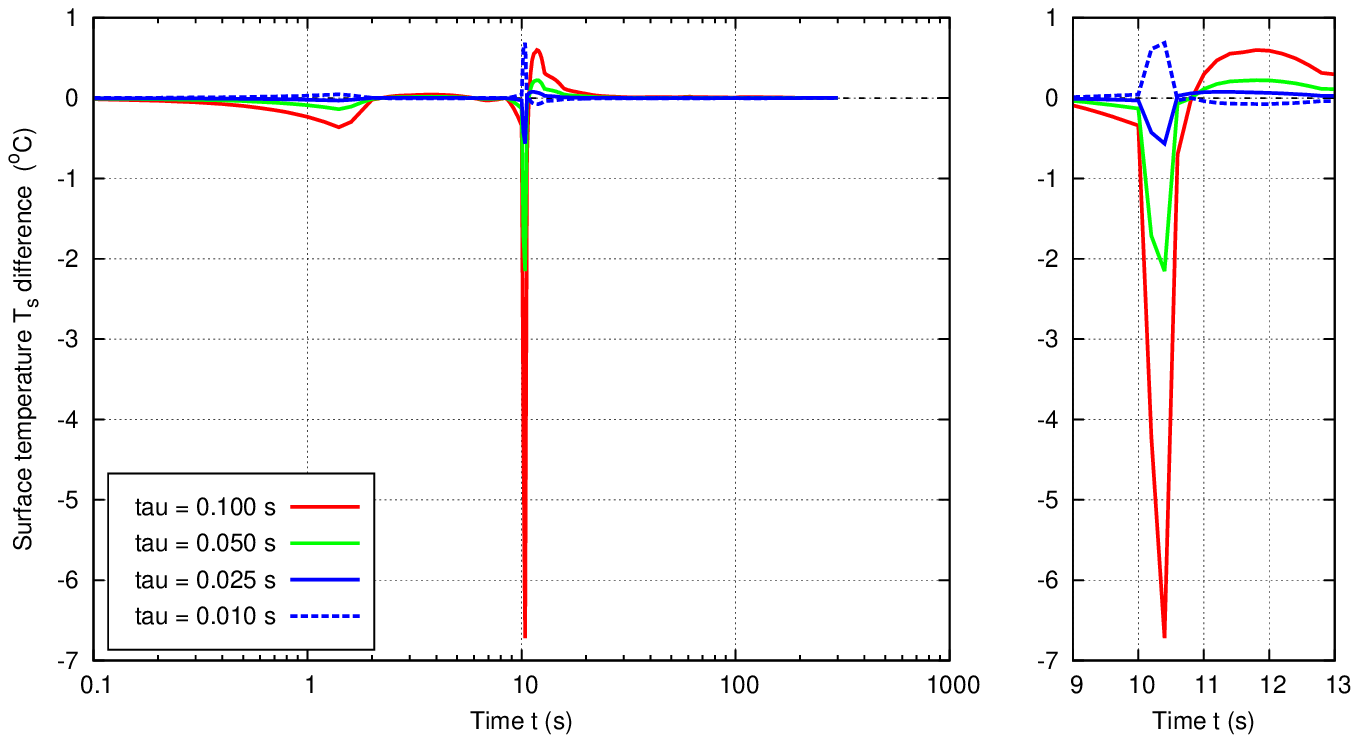}
\end{center}%
\Fig{Wat_Tau_Ts}%

\vspace*{-3pt}
\noindent
\hbox{Fig.~\ref{Wat_Tau_Ts} \ }%
	Water (time step variation):
	surface temperature $T_s$ differences.
\vspace*{12pt}
%

\begin{center}
	\epsfbox{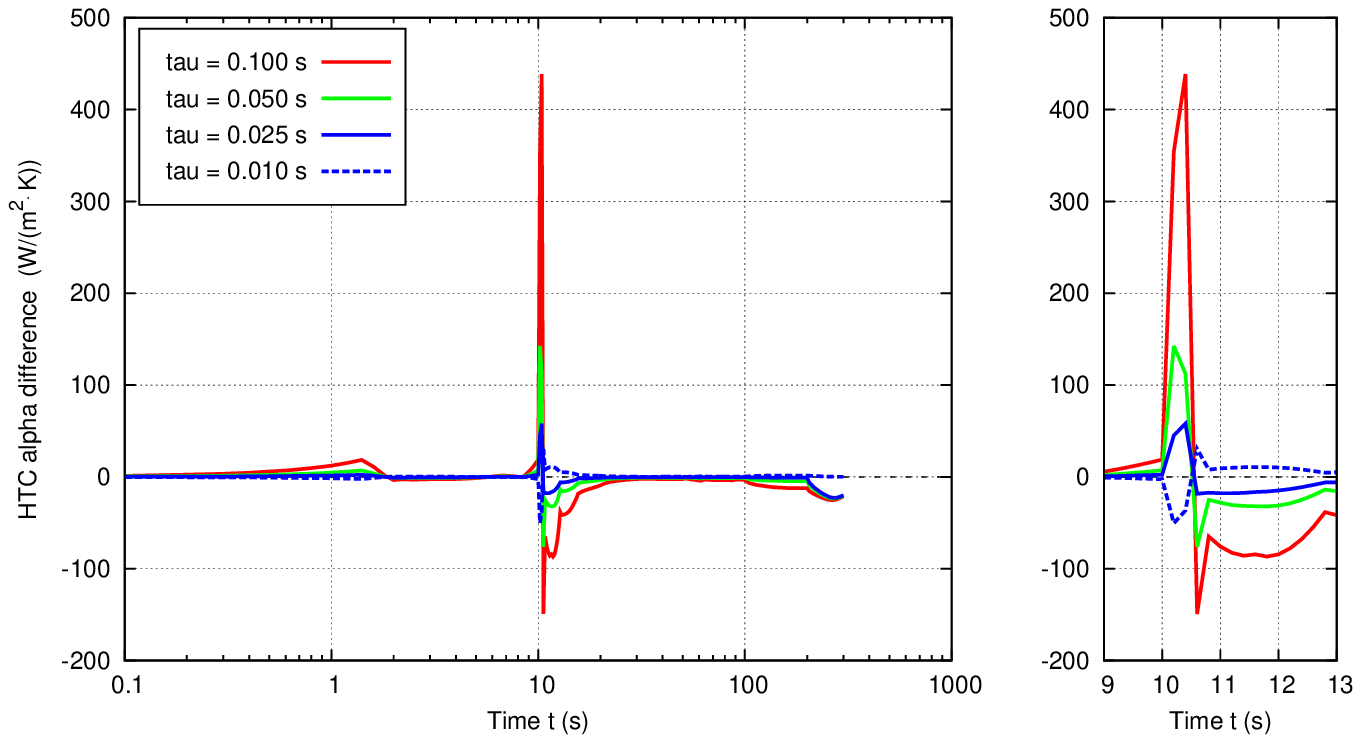}
\end{center}%
\Fig{Wat_Tau_Alpha}%

\vspace*{-3pt}
\noindent
\hbox{Fig.~\ref{Wat_Tau_Alpha} \ }%
	Water (time step variation):
	HTC $\alpha$ differences.
\vspace*{12pt}
%

\begin{center}
	\epsfbox{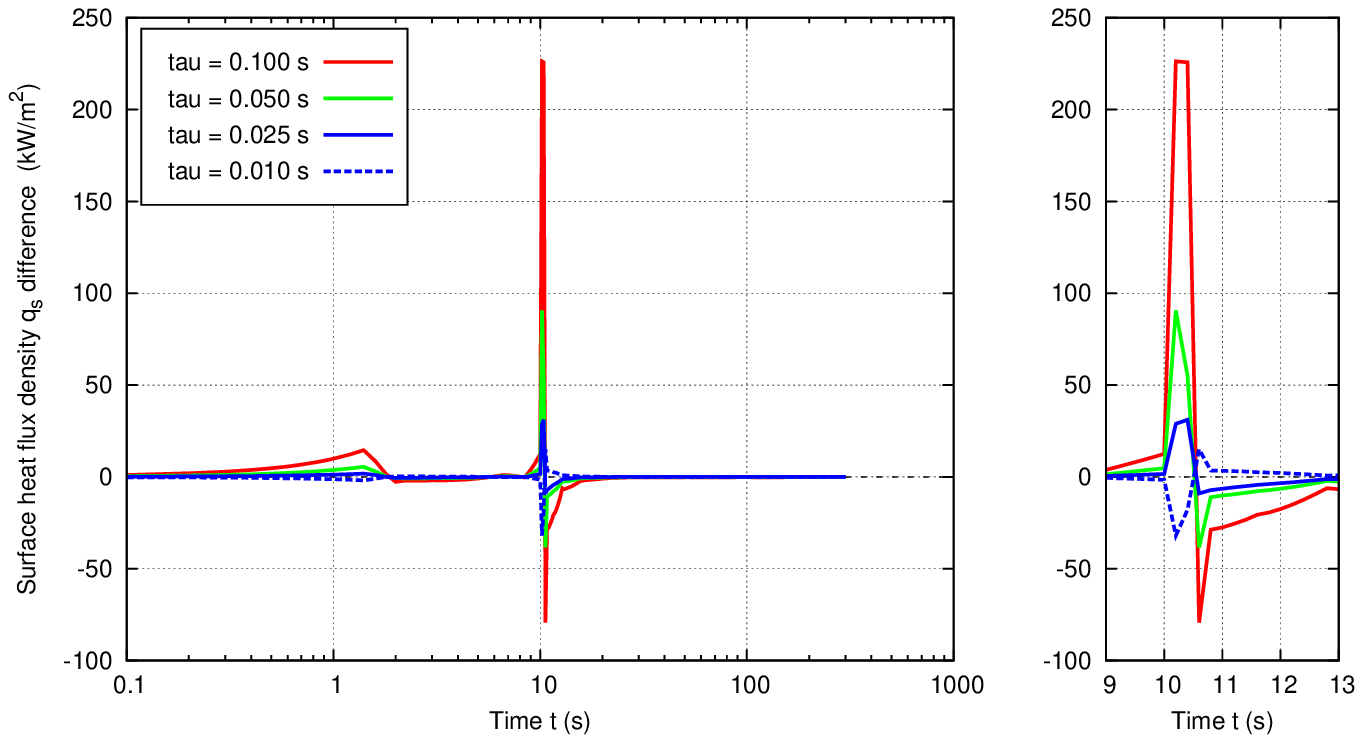}
\end{center}%
\Fig{Wat_Tau_Flux}%

\vspace*{-3pt}
\noindent
\hbox{Fig.~\ref{Wat_Tau_Flux} \ }%
	Water (time step variation):
	surface heat flux density $q_s$ differences.
\vspace*{12pt}
%

	Again, high differences in the HTCs are strictly confined to
the intensive cooling period. However, for the largest time step, the peak
differences are equivalent to the relative error of about $5 \%$
in the corresponding peak HTC values. Therefore, time steps much larger than
the standard step should not be used for the HTC calculation.

  	During the course of these tests, several other methods for
the solution of the IHCP have been tested, as well. For example, the finite
difference method \cite{Mitchell-Griffiths-1980} gives very similar results,
with the same discretization parameters, provided that an accurate algorithm
for numerical differentiation is used to calculate the surface heat flux
density.

%
%
\section{Conclusion}
%
%
	Several numerical experiments have been performed to test
the sensitivity of the HTC calculation with respect to the main input
parameters of the problem, including the numerical discretization parameters
involved in numerical solution of the problem.

	The results confirm that the numerical procedure itself is reliable
and numerically stable, so it can be used for the solution of the IHCP
to obtain reasonably accurate HTC curves, from properly prepared data.
The measured temperature curves have to be adequately smoothed, and
thermal properties of the material must be temperature-dependent to obtain
good results.

	All sensitivity tests have been conducted with realistic cooling
curves from two quenchants, and the results with respect to the position
of the near-surface thermocouple, and the diameter of the probe, may provide
some interesting information about the behavior of the HTC in various
liquid quenchants.
%
%

\newpage
%
%
%
%
\renewcommand{\refname}{References (in the order of citation)}

%
%
\end{document}